\documentclass[journal=nalefd,manuscript=article]{achemso}

\usepackage[T1]{fontenc}       
\usepackage{amsmath}
\usepackage{amssymb}

\newcommand*{\mybox}[1]{%
  \framebox{\raisebox{0pt}[0.4\baselineskip][0.05\baselineskip]{%
    #1}}}

\newcommand\ddfrac[2]{\frac{\displaystyle #1}{\displaystyle #2}}

\usepackage{color}
\definecolor{darkblue}{rgb}{0.2,0.2,0.7} 
\definecolor{darkred}{RGB}{139,0,0}

\usepackage{ragged2e}
\usepackage[bookmarks=true,pdfborder={0 0 0}]{hyperref}
\hypersetup{
  linkcolor  = darkblue,
  citecolor  = darkblue, 
  urlcolor   = darkblue, 
  colorlinks = true,
}

\makeatletter
\renewcommand*{\@cite@ofmt}{\hfil\color{red}}
\makeatother
   
\author{Clarice D. Aiello}
\affiliation[Chem]
{Department of Chemistry, University of California at Berkeley}
\alsoaffiliation[Stanford]
{Department of Bioengineering, Stanford University (present address)}
\email{caiello@stanford.edu}
\author{Andrea D. Pickel}
\affiliation[Mech]
{Department of Mechanical Engineering, University of California at Berkeley}
\author{Edward Barnard}
\affiliation[MF]
{Molecular Foundry, Lawrence Berkeley National Laboratory}
\author{Rebecca B. Wai}
\affiliation[Chem]
{Department of Chemistry, University of California at Berkeley}
\author{Christian Monachon}
\affiliation[Atto]
{Attolight AG, EPFL Innovation Park D, 1015 Lausanne, Switzerland}
\author{Edward Wong}
\affiliation[MF]
{Molecular Foundry, Lawrence Berkeley National Laboratory}
\author{Shaul Aloni}
\affiliation[MF]
{Molecular Foundry, Lawrence Berkeley National Laboratory}
\author{D. Frank Ogletree}
\affiliation[MF]
{Molecular Foundry, Lawrence Berkeley National Laboratory}
\author{Chris Dames}
\affiliation[Mech]
{Department of Mechanical Engineering, University of California at Berkeley}
\author{Naomi Ginsberg}
\affiliation[Chem]
{Department of Chemistry, University of California at Berkeley}
\alsoaffiliation[Phys]
{Department of Physics, University of California at Berkeley}
\alsoaffiliation[Biop]
{Molecular Biophysics and Integrative Bioimaging Division, Lawrence Berkeley National Laboratory}
\alsoaffiliation[Kav]
{Kavli Energy NanoScience Institute, Berkeley}

\title{Cathodoluminescence-based \\ nanoscopic thermometry \\ in a lanthanide-doped phosphor}

\begin{document}

\clearpage

Crucial to analyze phenomena as varied as plasmonic hot spots and the spread of cancer in living tissue, nanoscale thermometry is challenging: probes are usually larger than the sample under study, and contact techniques may alter the sample temperature itself. Many photostable nanomaterials whose luminescence is temperature-dependent, such as lanthanide-doped phosphors, have been shown to be good non-contact thermometric sensors when optically excited. Using such nanomaterials, in this work we accomplished the key milestone of enabling far-field thermometry with a spatial resolution that is not diffraction-limited at readout.

We explore thermal effects on the cathodoluminescence of lanthanide-doped NaYF$_4$ nanoparticles. Whereas cathodoluminescence from such lanthanide-doped nanomaterials has been previously observed, here we use quantitative features of such emission for the first time towards an application beyond localization. We demonstrate a thermometry scheme that is based on cathodoluminescence lifetime changes as a function of temperature that achieves $\sim$ 30 mK sensitivity in sub-$\mu$m nanoparticle patches. The scheme is robust against spurious effects related to electron beam radiation damage and optical alignment fluctuations.

We foresee the potential of single nanoparticles, of sheets of nanoparticles, and also of thin films of lanthanide-doped NaYF$_4$ to yield temperature information via cathodoluminescence changes when in the vicinity of a sample of interest; the phosphor may even protect the sample from direct contact to damaging electron beam radiation. Cathodoluminescence-based thermometry is thus a valuable novel tool towards temperature monitoring at the nanoscale, with broad applications including heat dissipation in miniaturized electronics and biological diagnostics. 


\clearpage

\vspace*{0.5cm}
\begin{figure}
  \centering
  \hspace*{-2cm}
  \includegraphics[width=0.65\textwidth]{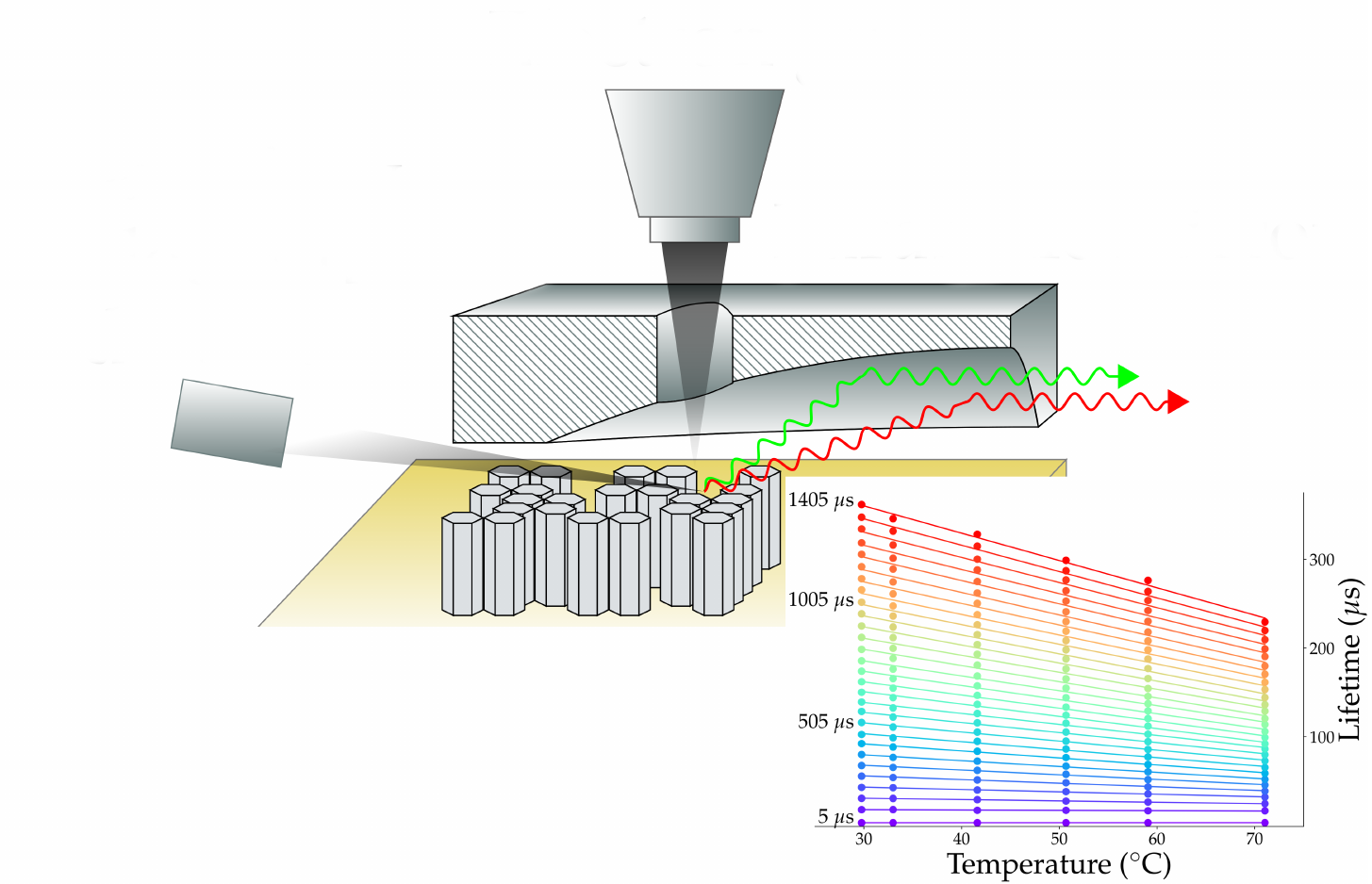}
\end{figure}
\vspace*{-1cm}
\section{Keywords}

Thermometry, cathodoluminescence, scanning electron microscopy, \newline lanthanide-doped nanoparticles

\clearpage

\section{Introduction}


Measuring temperature changes over nanoscopic distances in solid-state devices and in biological specimens is an outstanding challenge; local and minute temperature variations can yield information on the workings of miniaturized integrated circuits and plasmonic nanostructures, as well as marking the onset of cellular events. Monitoring and modeling these events is currently limited because, for such nanoscopic samples, contact-based thermometry is not suitable:\ the samples are themselves usually smaller than -- and potentially perturbed by -- the probe; even if nanotips are employed, the unknown thermal contact resistance cannot in general be guaranteed to be negligible when compared to the resistance between the tip and its thermal bath \cite{Gotsmann}. There is thus a compelling need for a new technique that provides non-contact, precise temperature and temperature change readings in microscopic samples.

Solid-state nanoemitters attract much attention given their potential for high-precision sensing at the nanoscale. In particular, lanthanide-doped nanophosphors are robust sensors, owing to a combination of highly desirable properties: sharp spectral lines \cite{review}, engineering tunability \cite{Chan12}, photo stability up to 600 K \cite{TM1} and biocompatibility \cite{nanoprobe, nanoprobe2, Diaz}.

Thermometry protocols based on lanthanide-doped nanomaterials usually rely on the stable change, as a function of temperature, of emitted light intensity at certain frequencies following optical excitation of the nanophosphor. Such protocols have known a substantial success \cite{TM1, TM2, TM3, TM4, NP1}, even if the spatial resolution of the yielded temperature information is capped by the diffraction limit. Whereas far-field super-resolution optical techniques (such as photo-activated localization, stochastic optical reconstruction and stimulated emission depletion microscopies \cite{superres1, superres2}) can overcome these issues, some of them cannot be used in a point-scanning fashion. Near-field scanning probe techniques such as near-field scanning optical microscopy \cite{snom1, snom2} provide sub-diffraction-limited resolution, but require complicated scanning hardware and are generally much slower. 

Here, we implement proof-of-principle experiments for a cathodoluminescence-based thermometry scheme employing lanthanide-doped nanophosphors. Upon excitation by a nanoscopic scanning electron beam (e-beam), NaYF$_4$: Yb$^{3+}$, Er$^{3+}$ nanoparticles emit cathodoluminescence, which can be collected and quantitatively analyzed with a time resolution down to 40 ns, and a nanometric spatial resolution limited by the electron scattering volume inside the material. We investigate two modalities of cathodoluminescence-based thermometry, and find that temperature information can be extracted from both light emission intensity and excited state lifetimes; in particular, excited state lifetime information can yield a sensitivity of $\sim$ 30 mK. Moreover, the scanning has minute timescales (a sub-$\mu$m patch can yield useful temperature information in $\sim$ 10 min), with single-pixel exposure to the e-beam of $\lesssim$ 500 $\mu$s (or shorter in the case of lifetime-based thermometry). Finally, we show that the spatial resolution of the method is limited only by the size of the electron scattering volume inside the material, which excites cathodoluminescence away from the e-beam impinging point. With the current experimentally accessible e-beam energies (10 keV), this resolution is $\sim$ 750 nm, approximately comparable to the resolution yielded by excitation with an optical, diffraction-limited scanning Gaussian beam; in the optical case, however, we estimate that a factor of 3 more photons are excited from a neighboring region and not from a target nanoparticle if compared with the present cathodoluminescence results. With an improved photon collection apparatus, enabling the use of lower e-beam energies (and concomitantly smaller scattering volumes), we estimate that this resolution would be comparable to the nanoparticles' size of $\sim$ 50 nm. On the other hand, the presently demonstrated cathodoluminescence edge sharpness is already nanoscopic at $\sim$ 30 nm. 

Combining the extensively researched lanthanide-doped nanophosphors with our novel thermometry scheme is a first step towards cathodoluminescent temperature mapping at the nanoscale. This technique adds to recent developments in non-invasive, not diffraction-limited cathodoluminescence-based sensors \cite{coconnor, coconnor2}.

\clearpage

\section{Experimental setup}

NaYF$_4$ co-doped with Yb$^{3+}$ and Er$^{3+}$ is a nanomaterial known for its upconversion properties under resonant optical excitation \cite{theo, chu,theo2, theo3, Chan12}. The Yb$^{3+}$ sensitizer ions absorb two photons at 980 nm, after which energy transfer to the Er$^{3+}$ emitter takes place, resulting in luminescence at shorter wavelengths. In this work, we quantitatively study the cathodoluminescence of hexagonal rods of NaYF$_4$ nanoparticles doped with 20$\%$ Yb$^{3+}$ and 2$\%$ Er$^{3+}$. In a simplified analogy, cathodoluminescence emission can be understood as if arising from an `incoherent excitation by a super-continuum source': high-energy electrons, as they penetrate and scatter through the material, create phonon-mediated excitations at high- and low-energetic levels alike \cite{ABAJO}. Whereas cathodoluminescence from such lanthanide-doped nanomaterials has been previously observed \cite{TM2, sinha}, here we characterize such emission quantitatively, with applications beyond localization.

\begin{figure}[h!]
\vspace*{-1.25cm}
  \centering
  \includegraphics[width=1\textwidth]{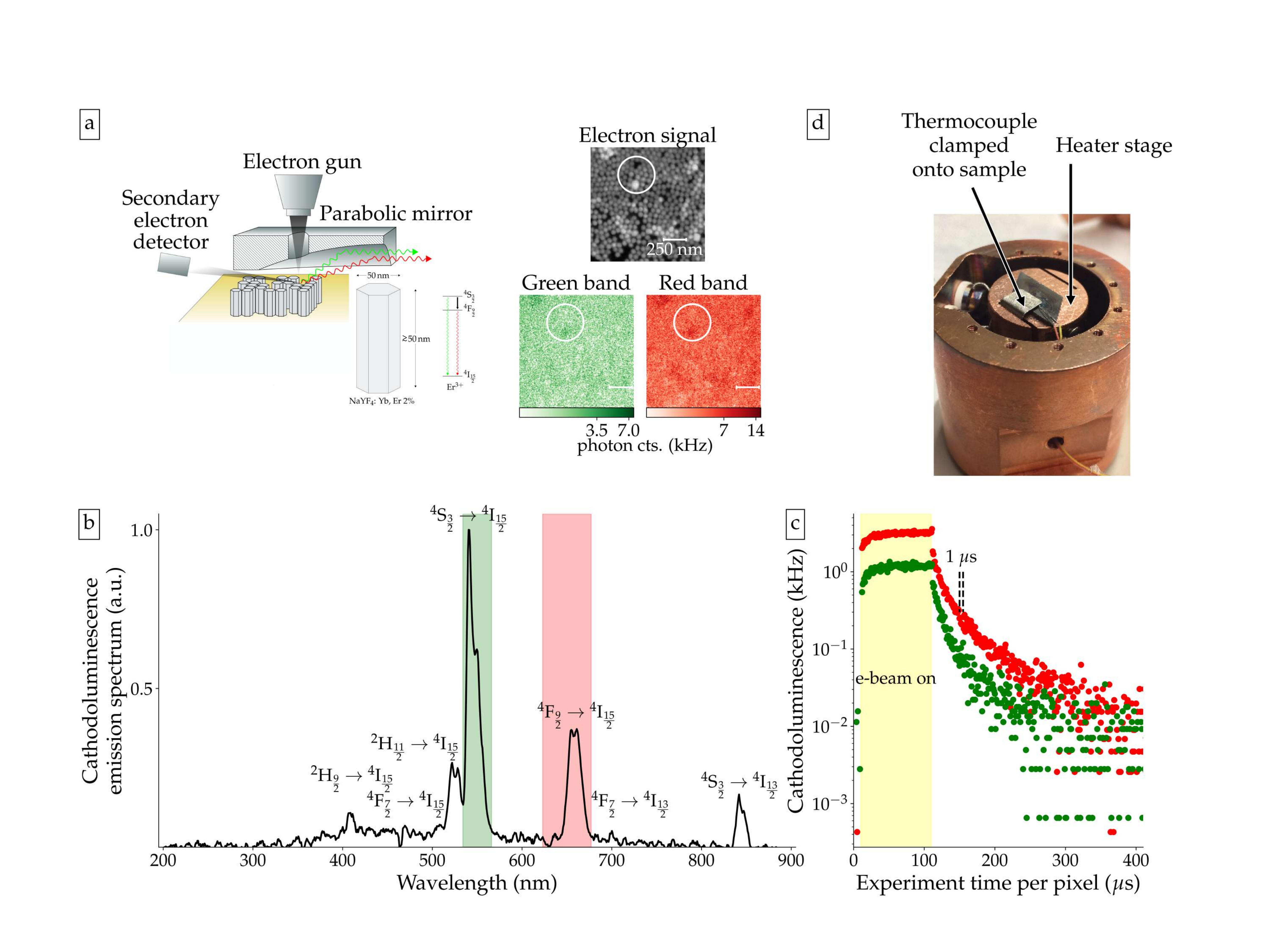}
  \caption{\textbf{Changes in NaYF$_4$: Yb$^{3+}$, Er$^{3+}$ nanoparticles' cathodoluminescence are studied as a function of temperature.} \mybox{\textbf{a}} Hexagonal rods of NaYF$_4$: Yb$^{3+}$, Er$^{3+}$ (50 nm diameter, $\gtrsim$ 50 nm height, deposited on silicon oxide on a silicon chip) are excited by a scanning electron beam. At every pixel of size 2.48 nm, the secondary electron signal is concomitantly recorded with the cathodoluminescence at different light bands by two photon multiplier tubes. The bands are here named green and red; these correspond to the Er$^{3+}$ transitions $^4$S$_{\frac{3}{2}}$ $\rightarrow$ $^4$I$_{\frac{15}{2}}$ and $^4$F$_{\frac{9}{2}}$ $\rightarrow$ $^4$I$_{\frac{15}{2}}$, respectively. Comparing the electron with the photon signals after a small patch is probed, we note with the help of the drawn white circle that, even though the nanoparticles are not clearly distinguishable, cathodoluminescence is locally modulated: regions without nanoparticles have lower cathodoluminescence and, similarly, particularly bright patches are also observed (the latter appear plotted as white patches under electron excitation).  Scale bars represent 250 nm. \mybox{\textbf{b}} At room temperature, the full cathodoluminescence spectrum of NaYF$_4$: Yb$^{3+}$, Er$^{3+}$ was also obtained using a spectrometer. Observed and labeled peaks correspond to Er$^{3+}$ transitions. Shaded green and red areas correspond to the bandpass filters (550 $\pm$ 16 nm; 650 $\pm$ 28 nm) used to determine the green and red light bands. \mybox{\textbf{c}} Quantitative cathodoluminescence statistics can be acquired per pixel (see supplementary information), under both steady and transient electron beam excitation conditions. A fast electrostatic beam blanker, with $\sim$ 25 ns rise time, enables the study of cathodoluminescence lifetime decays with time resolution down to 40 ns (here, a time resolution of 1 $\mu$s was chosen). The shown cathodoluminescence curves are averages over all of the 2.48 nm pixels in a 0.53 $\mu$m$^2$ nanoparticle patch, with 5 frame repetitions. \mybox{\textbf{d}} The effect of temperature on the green and red band transitions in Er$^{3+}$ is probed by placing the nanoparticle sample on a vacuum-compatible heater stage. A thermocouple is clamped onto the sample to monitor the temperature.}
  \label{Fig1}
\end{figure}

NaYF$_4$ co-doped with Yb$^{3+}$ and Er$^{3+}$ has been demonstrated to be photostable up to 600 K (up to 900 K in core-shell configurations) \cite{TM1} and biocompatible \cite{nanoprobe, nanoprobe2, Diaz}. As depicted in Fig.\ \ref{Fig1}\mybox{\textbf{a}}, cathodoluminescence from such nanoparticles (a monolayer of hexagonal rods, nominally of 50 nm diameter, $\gtrsim$ 50 nm height; drop cast on a silicon chip with a thin silicon oxide layer $\sim$ 200 nm) is excited by the e-beam of a field-emission scanning electron microscope (SEM). All experiments shown here were performed with an e-beam energy of 10 keV, and a current of 379 pA through a 30 $\mu$m aperture.

The SEM is equipped with a custom-built parabolic mirror \cite{Kaz} so that the resulting cathodoluminescence is transmitted though an optical window, allowing collection of photons outside the SEM vacuum chamber. Electron and cathodoluminescence signals are acquired pixel-wise at the same time, and can be used to correlate light emission with topography, the latter obtained using the secondary electron (SE) detector. 

A fast electrostatic beam blanker (rise time $\lesssim$ 25 ns) is synchronized to the scanning beam, such that pixel-wise cathodoluminescence can be recorded both under continuous excitation, and immediately after the e-beam is shut off, yielding cathodoluminescence lifetime information (minimum photon-collection integration of 40 ns, limited by the data acquisition card). The approximate instrument response function is a delta function on the ns scale (see Suppl.\ fig.\ 1\mybox{\textbf{c}}). 
Importantly, collection of electron and photon signals is concomitant; this means that the cathodoluminescence can be traced back to a region centered around individual nanoparticles as the secondary electron signal image is used to correlate light emission with topography.

By connecting a spectrometer to the optical port, we confirm that the cathodoluminescence emission spectrum, similarly to the fluorescence under optical excitation, is mainly due to the Er$^{3+}$ ion transitions, two of which we single out for the subsequent experiments (Fig.\ \ref{Fig1}\mybox{\textbf{b}}): $^4$S$_{\frac{3}{2}}$ $\rightarrow$ $^4$I$_{\frac{15}{2}}$ and $^4$F$_{\frac{9}{2}}$ $\rightarrow$ $^4$I$_{\frac{15}{2}}$, respectively constituting light bands separated by $\sim$ 0.4 eV. The sensitivity of the spectrometer is not sufficient to acquire data from sub-$\mu$m patches, and as such it is not used further in the experiments presented here, which rely exclusively on the use of more sensitive photon multiplier tubes (PMTs). Bandpass filters placed in front of the PMTs (passbands of 550 $\pm$ 16 nm and 650 $\pm$ 28 nm, indicated by the shaded areas in the figure) ensure that the collected cathodoluminescence is restricted to these bands. These 550 $\pm$ 16 nm and 650 $\pm$ 28 nm light bands are henceforth referred to as green and red wavelength bands. Given the collection efficiency of our setup, both bands produce cathodoluminescence counts in the kHz range when continuously excited. Even if the nanoparticles' topography under cathodoluminescence is less clearly distinguishable than the electron signal, as seen in Fig.\ \ref{Fig1}\mybox{\textbf{a}}, local modulations in the cathodoluminescence signal can be observed, for example, inside the white circle, which encompasses both a region devoid of nanoparticles, and one in which a NaYF$_4$: Yb$^{3+}$, Er$^{3+}$ nanomaterial clump cathodoluminesces at a higher rate than the material in its vicinity. 
  
We investigate the possibility of implementing nanoscopic thermometry using temperature-dependent changes in the cathodoluminescence of NaYF$_4$: Yb$^{3+}$, Er$^{3+}$. In order to do so, we place the sample onto the open-loop, resistively-heated stage of Fig.\ \ref{Fig1}\mybox{\textbf{d}}, which operates inside the SEM vacuum chamber.

In what follows, our work analyzes two features of the nanoparticles' cathodoluminescence as a function of temperature: its mean intensity and its excited-state lifetime. We first establish a baseline of temperature sensitivity using an intensity-based method; and subsequently improve upon it by monitoring cathodoluminescence lifetime information. With the latter method, we demonstrate $\sim$ 30 mK temperature sensitivity in a sub-$\mu$m patch.

\clearpage

\clearpage

\section{Temperature information is encoded \newline in the cathodoluminescence intensity}

\begin{figure}[h!]
  \centering
  \includegraphics[width=1\textwidth]{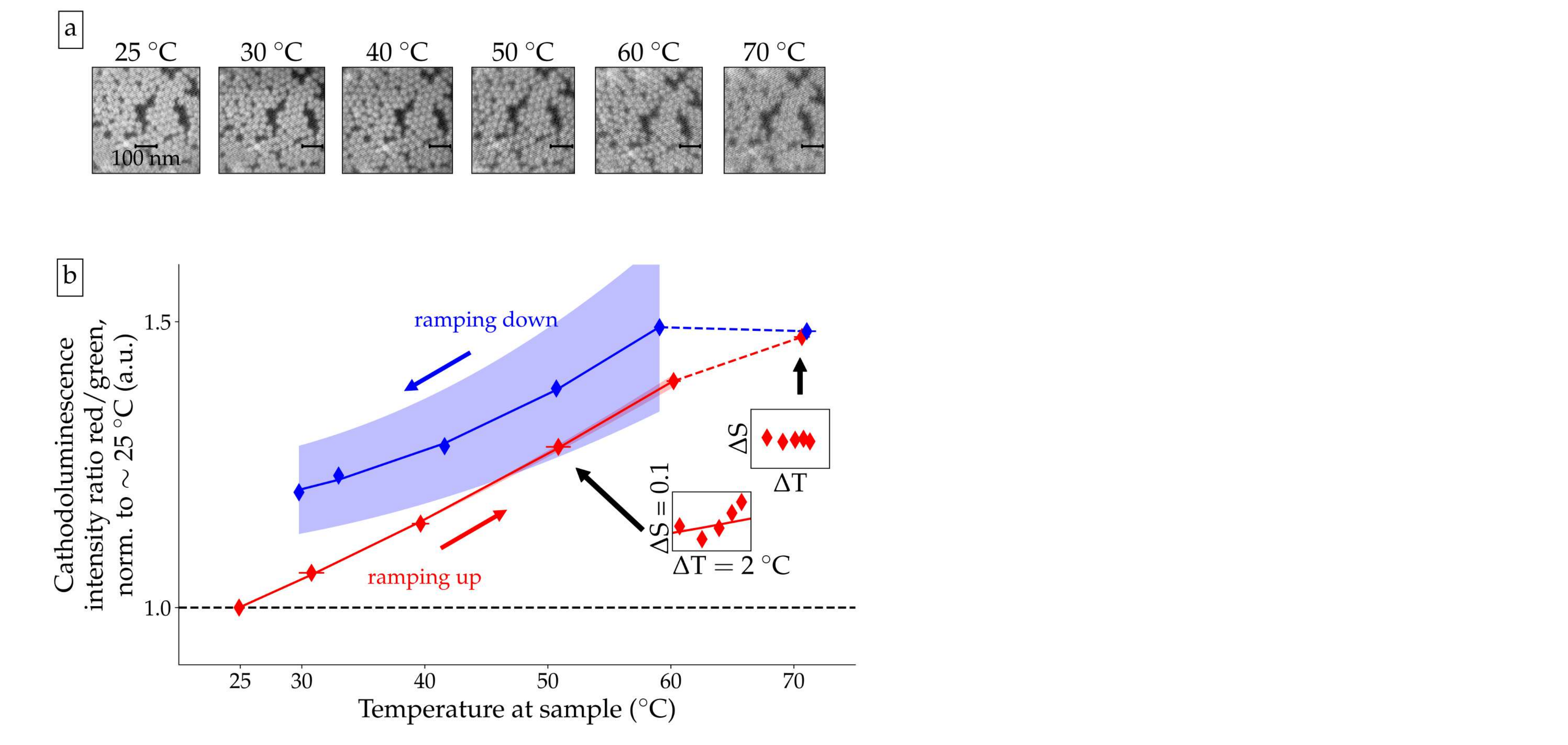}
  \caption{\textbf{Temperature information is encoded in the cathodoluminescence mean intensity ratio of sub-$\mu$m nanoparticle patches}; this baseline will be improved upon using cathodoluminescence lifetimes (Fig.\ \ref{Fig3}). \mybox{\textbf{a}} Electron signal from the same probed nanoparticle patch ($\sim$ 0.56 $\mu$m$^{2}$) at different sample temperatures; measurements, which take $\sim$ 10 min, are spaced in time by approximately 1 hour. \mybox{\textbf{b}} The cathodoluminescence ratios change by ramping up and down (blue and red curves) the sample temperature in steps of $\sim$ 10 $^{\circ}$C per hour (one temperature standard deviation marked by horizontal error bars), even with estimated thermalization times $\sim$ ms. We attribute this hysteresis to nanoparticle damage caused by continuous e-beam irradiation. 
Shown fits are to parabolas (excluding the $\sim$ 70 $^{\circ}$C data point), and the shaded areas represent the error in the model at the level of one standard deviation.}
  \label{Fig2}
\end{figure}

We first investigate thermal effects in the mean intensity of cathodoluminescence in the red and green light bands. Certain phonon-dependent transitions in NaYF$_4$: Yb$^{3+}$, Er$^{3+}$ are known to be affected by temperature changes \cite{Chan12,theo,fischer,sedl}, such that the intensity ratios of the luminescence signal of different spectral peaks can be used as a thermometer; this same effect is observed under cathodoluminescence in Fig.\ \ref{Fig2}. 

As measured in Suppl.\ fig.\ 5 \mybox{\textbf{b}}, even if the $^4$F$_{\frac{9}{2}}$ $\rightarrow$ $^4$I$_{\frac{15}{2}}$ transition rate is affected by temperature changes, the $^4$S$_{\frac{3}{2}}$ $\rightarrow$ $^4$I$_{\frac{15}{2}}$ transition has significantly larger rates at higher temperatures. The ratio of the cathodoluminescence from these two bands is stable (see Suppl. fig. 3), being independent of electron imaging parameters such as pixel size and e-beam current, and linearly depending on e-beam energy (see Suppl. fig. 4). In addition, such a ratiometric quantity does not depend on variations in nanoparticle coverage density. 

The e-beam continuously excites the same nanoparticle patch (0.56 $\mu$m$^{2}$, with nominal nanoparticle coverage $\sim$ 40.2 $\pm$ 6.8$\%$) for 150 $\mu$s per 2.48 nm pixel; the mean cathodoluminescence ratio is recorded for the entire time the e-beam is un-blanked. Results over the full field-of-view are averaged, and 5 immediately consecutive frames are taken at each temperature; the total data acquisition time (not counting computer processing time) at each temperature is 12 minutes. The measured temperature at each frame is automatically recorded and averaged over to yield one data point. The temperature is ramped up and down at a rate of $\sim$ 10 $^{\circ}$C per hour. The secondary electron signal, registered over the 5 frames, is depicted at different temperatures in Fig.\ \ref{Fig2}\mybox{\textbf{a}}; re-focusing and re-centering of the SE image was performed at every temperature step; we attribute the increasing blurriness at higher temperatures to thermally-induced mechanical drifts of the sample on the heater stage. The shown micrographs are slightly different in size because we choose to analyze only the regions that are common to all 5 frames at a given temperature. 

We expect the ratio of red to green cathodoluminescence to increase for increasing temperatures (a similar, spectrally-resolved result at a lower magnification is found in Suppl.\ fig.\ 5); this is due to the fact that the red $^4$F$_{\frac{9}{2}}$ $\rightarrow$ $^4$I$_{\frac{15}{2}}$ Er$^{+3}$ transition is more strongly phonon-coupled than the green $^4$S$_{\frac{3}{2}}$ $\rightarrow$ $^4$I$_{\frac{15}{2}}$ \cite{Chan12}. 

The cathodoluminescence intensity ratio is plotted in Fig.\ \ref{Fig2}\mybox{\textbf{b}} as a function of the temperature (temperature uncertainties at one standard deviation are plotted by horizontal error bars), after being normalized by the initial ratio at $\sim$ 25 $^{\circ}$C. This normalization reduces the effects of inevitably differing PMT alignments from day to day, and of differing PMT gains at the two considered wavelength bands. When increasing the sample temperature from room-temperature, the cathodoluminescence ratio increases parabolically (higher F-statistic than a linear model). Ramping the temperature back down leads also to a parabolic (same F-statistic metric) decrease of the cathodoluminescence ratio. A hysteresis emerges, also evident in the ratio values measured at the turning point at $\sim$ 70 $^{\circ}$C. 

The insets for individual frame averages at $\sim$ 50 and 70 $^{\circ}$C reveal that the cathodoluminescence intensity ratio ($\Delta S$) varies more steeply for a 2 $^{\circ}$C temperature change at the lower temperature. We fit parabolas ($a$ $\cdot$ ($T$ - 25 $^{\circ}$C)$^2$ $+$ $b$ $\cdot$ ($T$ - 25 $^{\circ}$C) $+$ $c$, with a fixed point $c$ $=$ 1 for the ramping up data set at increasing temperatures) to the cathodoluminescence ratios for temperatures ramping up and down; the error in the model is shown in the shaded areas (see Supplementary information for definition of error in the model).

This constitutes the first sub-$\mu$m quantitative demonstration of thermal features in the cathodoluminescence intensity of lanthanide-doped nanoparticles. The limitation of using cathodoluminescence intensity ratios to perform thermometry experiments is the nanoparticles' continuous exposure to the e-beam, which affects their mean emission in ways that could not be systematized and predicted by our experiments. Moreover, light collection efficiency fluctuations (e.g.\ PMTs relative alignment from day to day) would require that a calibration curve be taken at each thermometry measurement session, further increasing the nanothermometers' exposure to e-beam radiation. We address and overcome these concerns in the following, separate approach, which is demonstrated to be sensitive to thermal effects in the mK-range based on changes in cathodoluminescence excited-state lifetimes.

\clearpage

\clearpage

\section{Temperature information, \newline with a sensitivity down to $\sim$ 30 mK, \newline is encoded in the cathodoluminescence lifetimes}

An alternative scheme can provide advantages relative to cathodoluminescence intensity-based thermometry such as reduced electron beam exposure time and independence from alignment and collection efficiency variations if excited-state lifetimes can be shown to depend on temperature; we investigate if this hypothesis holds true under cathodoluminescence here. 
We demonstrate that cathodoluminescence lifetimes can be used to measure relative temperature changes down to the mK-range. 

Using a lifetime decay signal to perform thermometry is attractive in that the exposure of the nanothermometer to the damaging e-beam radiation is minimized. The nanoparticle needs only be excited long enough to stabilize the cathodoluminescence intensity, following a rising transient measured to be $\sim$ 30 $\mu$s, see Suppl.\ fig.\ 3\mybox{\textbf{d}}; this shorter excitation duration translates into a five-fold decrease in radiation dose as compared to the cathodoluminescence intensity-based thermal experiments of Fig.\ \ref{Fig2}. Moreover, such a (normalized) lifetime decay signal is independent of cathodoluminescence collection efficiency fluctuations and background level, which vary from day to day; of the local concentration of nanothermometers; and, importantly, on local particle damage owing to prolonged e-beam irradiation (assuming that damage is stochastic; and that only undamaged nanoparticle material contributes to the cathodoluminescence signal). Furthermore, and in contrast to intensity-based thermometry, in principle only one spectral line needs to be measured. 

\begin{figure}[h!]
  \centering
  \includegraphics[width=1\textwidth]{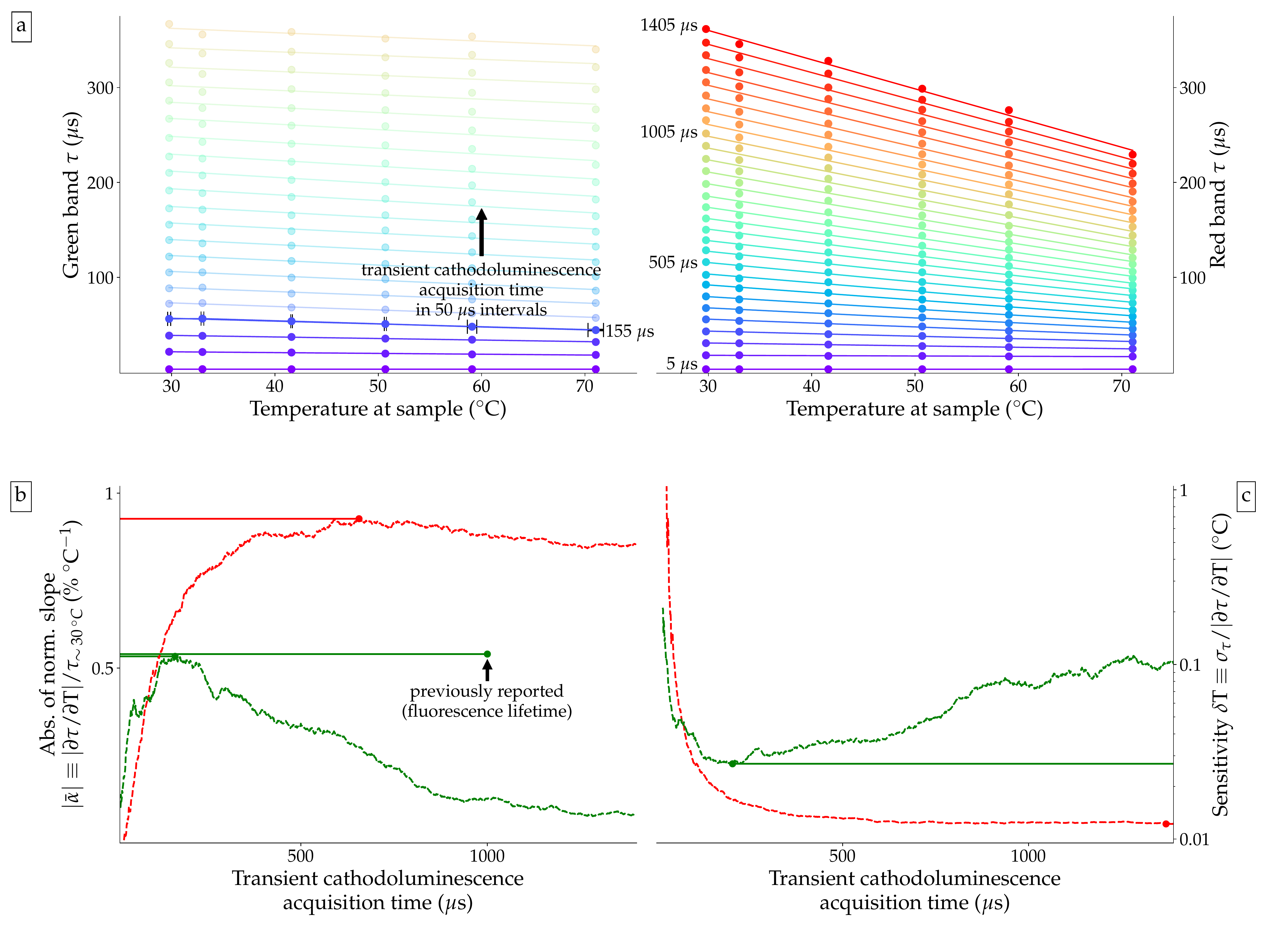}
  \caption{\textbf{Temperature information is encoded in the cathodoluminescence lifetimes, with a sensitivity down to 30 mK.} \mybox{\textbf{a}} The calculated characteristic decay times $\tau(t,T)$ are plotted for discrete and increasing (rainbow palette) transient cathodoluminescence acquisition times $t$, in intervals of 50 $\mu$s, for different temperatures of a nanoparticle patch being heated up over a 40 $^{\circ}$C range. Linear fits over the different temperatures are shown for each plotted $t$. The change of $\tau(t,T)$ with temperature is more pronounced for the red light band, which is more strongly phonon-coupled and has a better signal-to-noise figure in our setup. \mybox{\textbf{b}} The absolute value of the fitted linear slopes can be normalized and plotted as a function of increasing transient cathodoluminescence acquisition times. Those curves for $|\overline{\alpha}(t)|$ indicate a change in measured $\tau(t,T)$ as more transient data is being acquired; a large value of $|\overline{\alpha}(t)|$ is desirable. $|\overline{\alpha}(t)|$ for the red light band outperforms the green light band. \mybox{\textbf{c}} We calculate the optimal transient cathodoluminescence acquisition time by plotting and finding the minimum of the sensitivity, or noise-limited temperature resolution. For the green band, the sensitivity is minimized for times close to those that maximize $|\overline{\alpha}(t)|$; for the red band, the sensitivity is minimal at the end of the acquisition interval. For the red (green) color band, the sensitivity reaches $\sim$ 12 mK ($\sim$ 27 mK), which is a consequence of the fact that each point in  \mybox{\textbf{a}} is an average over 4.5 $\times$ 10$^5$ individual cathodoluminescence decay measurements.}
  \label{Fig3}
\end{figure}

The present measurement scheme analyses the cathodoluminescence decay, pixel-by-pixel, of a nanoparticle patch (0.55 $\mu$m$^{2}$, with nominal nanoparticle coverage $\sim$ 51.2 $\pm$ 4.8$\%$) at different temperatures, with 1 $\mu$s time resolution over 1400 $\mu$s following fast e-beam blanking ($\sim$ 25 ns); excitation conditions are identical to the cathodoluminescence intensity ratio measurements of Fig.\ \ref{Fig2}. The temperatures at the thermocouple in each frame are automatically recorded and averaged to yield a single value to represent each condition. The mean temperatures thus obtained are $\{71.05 \pm 0.75, 59.08 \pm 0.46, 50.69 \pm 0.10, 41.61 \pm 0.08, 32.99 \pm 0.13, 29.77 \pm 0.15 \}$ $^{\circ}$C, in the order at which the data were taken; temperature uncertainties are indicated by black horizontal error bars signaling one standard deviation on the topmost curve of the left plot in Fig.\ \ref{Fig3}\mybox{\textbf{a}}. 

We define a monotonically increasing characteristic lifetime decay \cite{Diaz}
\begin{equation}
\tau(t, T) \equiv \ddfrac{\int_{0}^{t} I(t',T)\cdot t' dt'}{\int_{0}^{t} I(t',T)dt'} \ ,
\end{equation}
where $I(t,T)$ is the cathodoluminescence decay intensity in either the red or the green light band, and $t$ is the time after beam blanking ($t_{\textrm{beam blanking}} = 0$). Making use of cumulative photon counts by treating the cathodoluminescence decay intensity as a probability distribution is equivalent to low-pass filtering, and is of importance as the signal-to-noise is progressively reduced upon the measurement of an exponentially decaying quantity.

In Fig.\ \ref{Fig3}\mybox{\textbf{a}}, we plot typical $\tau(t,T)$ for the green and red bands, as a function of nanoparticle temperature, for select transient cathodoluminescence acquisition times $t$; different marker colors signal different $t$, in 50 $\mu$s intervals, starting at 5 $\mu$s (i.e.,\ $\tau(t,T)$ are calculated at different temperatures, and plotted for $t = \{$5, 55, 105, 155, ...$\}$ $\mu$s following a rainbow palette). A linear fit ($\tau(t, T) = \alpha(t)\cdot T + \beta(t)$) at each $t$ interval is shown.

The lifetimes of the normalized $^4$S$_{\frac{3}{2}}$ $\rightarrow$ $^4$I$_{\frac{15}{2}}$ green band transition 
are linearly fit with slopes that approach zero. In turn, the normalized cathodoluminescence decay of the $^4$F$_{\frac{9}{2}}$ $\rightarrow$ $^4$I$_{\frac{15}{2}}$ red band transition accelerates with increasing temperature, as expected by a phonon-assisted increase in net de-excitation probability, in an approximately linear fashion for the measured range (also observed elsewhere \cite{Diaz}); this is apparent in the increasing slope of the linear fit at different $t$, shown up to 1405 $\mu$s. Previous optical studies have indeed found that decay lifetimes are longer for the red band transition \cite{theo2}.

In Fig.\ \ref{Fig3}\mybox{\textbf{b}}, the absolute value of the linear best-fit slope $|\alpha(t)|$, normalized by the characteristic lifetime taken at the lowest nanoparticle temperature of $\sim$ 30 $^{\circ}$C \cite{Diaz}, 
\begin{equation}
|\overline{\alpha}(t)| \equiv \ddfrac{\frac{\partial\tau(t, T)}{\partial T}}{\tau(t,\sim \textrm{30 } ^{o}\textrm{C})} \ ,
\end{equation}
is shown as a percentage per $^{\circ}$C, for increasing times after beam blanking. In other words, $|\overline{\alpha}(t)|$ quantifies the change in the apparent lifetime $\tau(t, T)$ that we measured over a 40 $^{\circ}$C range. The maximum $|\overline{\alpha}(t)|$ values for each light band are indicated by a circle (at $\sim$ 656 and 163 $\mu$s for the red and green bands, respectively, corresponding to 0.93 and 0.53$\%$ $^{\circ}$C$^{-1}$); also depicted is a previously reported $|\overline{\alpha}(t)| \sim$ 0.54$\%$  $^{\circ}$C$^{-1}$ for fluorescence lifetimes in the green band, estimated following the same method after 1000 $\mu$s of transient acquisition \cite{Diaz}.

From this plot, it is immediately clear that cathodoluminescence lifetime thermometry is more sensitive for strongly phonon-coupled transitions: analysis of the red band cathodoluminescence decay yields larger $|\overline{\alpha}(t)|$. Additionally, in the present experiments, raw cathodoluminescence counts (with e-beam on) in the red light band are 2--3.5 fold stronger than those in the green band; the exact figure depends on PMT sensitivity and alignment. Hence, the fact that $|\overline{\alpha}(t)|$ sharply \textit{decreases} for the green band after only $\sim$ 163 $\mu$s is both a consequence of less strong phonon coupling controlling the dynamics of the $^4$S$_{\frac{3}{2}}$ $\rightarrow$ $^4$I$_{\frac{15}{2}}$ transition; and of a decreased signal-to-noise ratio. The information in this plot, moreover, can be used to determine the optimal transient acquisition time, in order to achieve the most sensitive measurement. 

To determine the transient acquisition time that enables the most sensitive cathodoluminescence-based temperature measurement, we numerically calculate and plot in Fig.\ \ref{Fig3}\mybox{\textbf{c}} the measurement sensitivity in $^{\circ}$C, or noise-limited temperature resolution, defined as \cite{magneto, npbrazil} 
\begin{equation}
\partial T(t) \equiv \ddfrac{\sigma_{\tau}(t)}{\left|\frac{\partial\tau(t,T)}{\partial T}\right|} \ .
\end{equation}
Here, $\sigma_{\tau}$ is the standard deviation (uncertainty) of the fitted value for $\tau(t,T)$, and $|\partial\tau(t,T)/\partial T|$ is the unnormalized absolute value of the slope of the linear fit, or $|\alpha(t)|$ (details on the calculation of $\partial T$ are provided in the Supplementary Information). Importantly, $\partial T$ quantifies the magnitude of noise-related variations in the thermometry signal compared to the magnitude of changes in the parameter being measured. Effectively, $\partial T$ represents the \textit{minimal difference in temperature} that can be measured, given the available data, as a function of acquisition time. It is insensitive to the standard deviations of the measured sample temperature during calibration, even if these are substantially larger (horizontal errorbars in Fig.\ \ref{Fig3}\mybox{\textbf{a}}): the measurement of the sample temperature via a thermocouple provides a calibration, and as such need not be more sensitive than the temperature information yielded by cathodoluminescence-based lifetime thermometry. Moreover, $\partial T$ does \textit{not} inform how good the nanothermometer is in measuring absolute temperatures; rather, it points to the minimum difference in temperature that would be detected by our scheme, given the noise levels. In other words, it is a measurement of precision, not accuracy. In many applications, it is indeed more relevant to measure temperature differences (eg.,\ by how much a cancerous cell's temperature changes as a result of exothermic anaerobic processes \cite{cancer1, cancer2}) from a baseline than to measure absolute temperatures.

The optimal transient acquisition time for our cathodoluminescence-based lifetime thermometry is precisely the time that minimizes the sensitivity. In the present measurements, this optimal time is as short as 204 $\mu$s for the green light band, yielding $\partial T \sim$ 27 mK; accordingly, in Fig.\ \ref{Fig3}\mybox{\textbf{a}}, the calculated green band times $\tau(t,T)$ are plotted in 50 $\mu$s intervals up to times shorter than 204 $\mu$s ($\tau(t,T)$ curves for longer acquisition times have essentially the same slope, confirming their low thermometry information content; they are plotted with transparency). The acquisition time that minimizes the sensitivity matches within tens of $\mu$s the time for which $|\overline{\alpha}(t)|$ is at a maximum. Conversely, the optimal transient acquisition time is $\sim$ 1370 $\mu$s for the red light band, at which point sensitivity is at a minimum over the available data acquisition duration, for $\partial T \sim$ 12 mK. This mK-level sensitivity reflects the fact that each point in Fig.\ \ref{Fig3}\mybox{\textbf{a}} is an average over 4.5 $\times$ 10$^5$ individual cathodoluminescence decay measurements.  (A plot similar to Fig.\ \ref{Fig3}, for data taken at increasing temperatures, can be found in Suppl. fig. 6; best sensitivities are on the same order of magnitude.)

\clearpage

\clearpage

\section{Spatial resolution \newline of cathodoluminescence-based thermometry}

\begin{figure}[h!]
  \centering
  \includegraphics[width=1\textwidth]{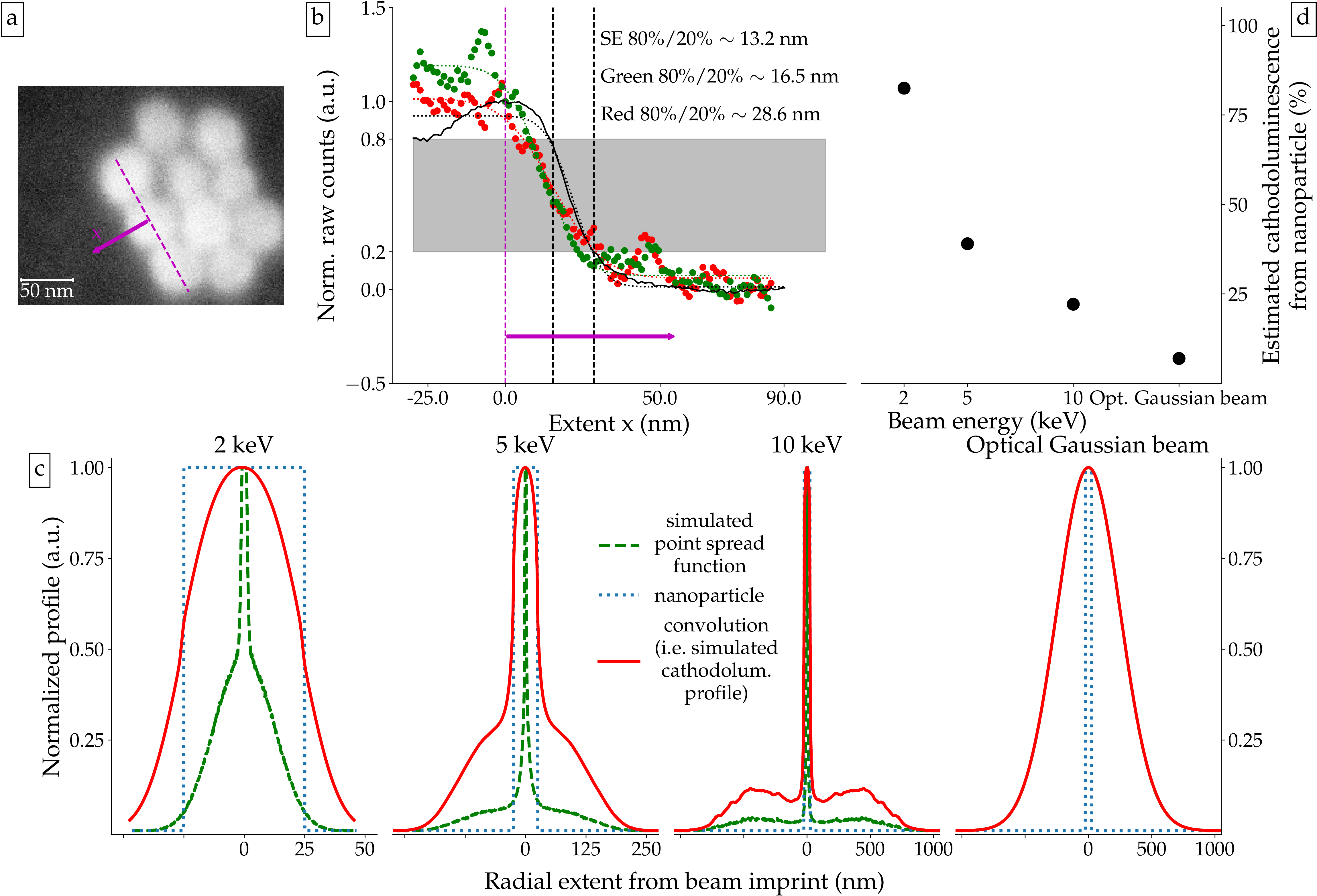}
  \caption{\textbf{The spatial resolution of cathodoluminescence-based thermometry outperforms that of optical methods.} \mybox{\textbf{a}} and \mybox{\textbf{b}} The  cathodoluminescence edge resolution, as estimated by the signal decay from the center of a row of 3 nanoparticles, is comparable to that of secondary electrons using both a 80$\%$/20$\%$ metric, and a least-squares fit (dotted lines) to an error function. \mybox{\textbf{c}} It is crucial to estimate the extent over which the electron beam excites phonon modes and cathodoluminescence, given the electron scattering inside the material. Numerical simulations of cathodoluminescence point spread functions for different electron beam energies reveal bi-modal distributions, with a `shoulder' which is as large as $\sim$ 750 nm for a 10 keV beam, but only roughly the spatial extent of one nanoparticle for a 2 keV beam. The resulting cathodoluminescence profiles limit the spatial resolution of nanoscopic thermometry by approximately the `shoulder' width. Nevertheless, in comparison, optical excitation methods have an even broader luminescence profile. \mybox{\textbf{d}} Given the electron scattering, if one nanoparticle in a sheet of contiguous nanoparticles were to be excited, the percentage of luminescence coming from the excited nanoparticle is drastically reduced from over 80$\%$ at 2 keV to only 22$\%$ at 10 keV; if the excitation were optical, this figure drops further to  7$\%$.}
  \label{Fig4}
\end{figure}

We quantify the spatial resolution of the cathodoluminescence intensity signal by comparing a feature's edge sharpness in cathodoluminescence intensity and in the SE signal using a 10 keV e-beam. This is depicted in Fig.\ \ref{Fig4}\mybox{\textbf{a}}: the dashed magenta line runs across the center of 3 nanoparticles; we quantify the steady-state luminescence and SE signal profiles along the perpendicular direction marked by the magenta arrow. Such profiles, depicted in Fig.\ \ref{Fig4}\mybox{\textbf{b}}, are obtained following a 150 $\mu$s-long excitation per 1.1 nm pixel, with the cathodoluminescence counts integrated over the complete excitation duration, for 10 frame averages; note that the micrograph and corresponding cathodoluminescence are acquired in less than 2 minutes. Raw photon counts and SE greyscale counts are normalized to 1 at the dashed magenta line; we estimate the edge resolution of the SE signal by the decay extent between 80$\%$ and 20$\%$ bands, obtaining $\sim$ 13.2 nm. We apply a 5-pixel ($\sim$ 5.5 nm) moving average to the cathodoluminescence signals, and similarly estimate the decay extent to be $\sim$ 16.5 nm and $\sim$ 28.6 nm for the green and red light bands, respectively. In addition, we fit the curves to an error function (dotted lines), yielding standard deviations for the associated Gaussian of 7.4 $\pm$ 0.2 nm, 9.9 $\pm$ 0.4 nm, 13.0 $\pm$ 0.9 nm for the SE, green and red curves, respectively; the values obtained by doubling these standard deviations are in good agreement with the 80$\%$ and 20$\%$ bands estimates. Both the fitted and the 80$\%$ and 20$\%$ width estimates indicate that the cathodoluminescence edge sharpness is on the same order of magnitude as the SE signal's. 

A relevant issue in the above-described thermometry scheme is quantifying how much of the cathodoluminescence signal is generated by exciting nanoparticles that are not colocalized with the primary impinging point of the electron beam, which can occur due to electron scattering inside the substrate material beneath the nanoparticles. This phenomenon, akin to the proximity effect thoroughly studied in the context of electron beam lithography \cite{pe1, pe2, pe3}, accounts for large-distance excitation of cathodoluminescence by scattered electrons, a diffuse background that degrades the intrinsic spatial resolution of cathodoluminescence-based thermometry.

We performed numerical Monte-Carlo simulations of inelastically scattered electron trajectories at different beam energies (details can be found in Suppl. fig. 7), yielding the simulated cathodoluminescence point spread functions plotted in dashed green in Fig.\ \ref{Fig4}\mybox{\textbf{c}}. For 2, 5 and 10 keV beam energies, the point spread function is bi-modal; it is composed of a very tight central peak the size of the incident beam ($\sim$ 2 nm), on top of a much weaker but broader diffuse background, with radii $\sim$ 20 nm, 150 nm and 750 nm, respectively. The simulated cathodoluminescence profile, depicted with the solid red line, is obtained after a one-dimensional convolution of the point spread function with a single nanoparticle profile, the latter depicted using the blue dotted line. Incidentally, the estimated cathodoluminescence edge sharpness of Fig.\ \ref{Fig4}\mybox{\textbf{b}} are on the same order of magnitude as half-width at half-maximum of the central distribution composing the 10 keV cathodoluminescence point spread function, $\sim$ 33 nm.

If a uniform sheet of contiguous nanoparticles were used as nanothermometers, it is straightforward to estimate the percentage of cathodoluminescence collected from the single nanoparticle (over its 50 nm of diameter) that is directly excited by the primary electron beam aligned with its center; as compared to the undesired cathodoluminescence due to scattered electrons triggering excitations from surrounding particles. As seen in Fig.\ \ref{Fig4}\mybox{\textbf{d}}, it is clear that, given the proximity effect, most luminescence intensity, even for beam energies as low as 5 keV, does not originate from the excited nanoparticle, but rather from its neighbors. At the 10 keV e-beam energy employed in our work, only 22$\%$ of the luminescence comes from the nanoparticle on which the impinging e-beam is focused. A cathodoluminescence-based thermometry signal would thus yield averaged spatial information over distances comparable to the width of the broader distribution of the point spread function (the cathodoluminescence intensity is weight-averaged by the convolutions depicted in Fig.\ \ref{Fig4}\mybox{\textbf{c}}), amounting to $\sim$ 750 nm for the e-beam energy of 10 keV used throughout the experiments. For an e-beam energy on the order of 2 keV, however, this resolution should approach the nanoparticles' size at $\sim$ 50 nm, and more than 80$\%$ of the cathodoluminescence would indeed come from the single excited particle. We provide in the Supp. fig. 7 a simulation of cathodoluminescence profiles for nearby nanoparticles excited by different e-beam energies. 

For the present experiments, limitations of photon collection efficiency and electron beam current tunability prevented useful signals for beam energies lower than 10 keV from being acquired. Yet, despite the compromise in spatial resolution of nanoscale thermometry in this study at 10 keV, and which could be eliminated through further technical improvements, the present method surpasses the resolution of more typical optical methods. As shown in Fig.\ \ref{Fig4}\mybox{\textbf{c}}, a Gaussian beam with a diffraction-limited waist taken to be 250 nm (full-width at half-maximum $\sim$ 589 nm) is simulated to excite a single nanoparticle, yielding a similarly broad luminescence profile; as estimated in Fig.\ \ref{Fig4}\mybox{\textbf{d}}, if a sheet of nanoparticles were to be excited by such a Gaussian beam, less than 7$\%$ of the luminescence would come from the particle at which the excitation is centered -- thus, the figure for the demonstrated cathodoluminescence method, at $\sim$ 22$\%$, is already more than three-fold superior. Increasing photon collection efficiency can at once improve the sensitivity (higher signal-to-noise ratios decrease $\partial T$) and the resolution (enabling the use of much lower e-beam energies) of cathodoluminescence-based thermometry.

For nanoscopic thermometry, we envision using a sheet of lanthanide-doped nanoparticles drop cast onto the sample of interest. If single or sparsely-distributed nanoparticles were used as nanothermometers, the spatial resolution of the method would automatically improve, as scattered electrons would not excite nearby phosphors nor contribute to a diffuse background. In this case, only the central peak of the calculated point spread functions is relevant, and the spatial resolution is exclusively limited by the nanometric diameter of the primary e-beam. There is consequently a trade-off between temperature information coverage and spatial resolution.

An alternative scheme would have a thin film of NaYF$_4$: Yb$^{3+}$, Er$^{3+}$ covering the sample of interest; the sample can be deposited or dropcast onto the thin film side which is not directly irradiated by the e-beam. The nanomaterial could be probed by the e-beam and convey temperature information, while the sample itself needs never be directly exposed to the e-beam radiation, much as in cathodoluminescence-activated imaging by resonant energy transfer (CLAIRE) \cite{coconnor, coconnor2}. In CLAIRE as in our proposed thermometry scheme, low e-beam energies are required to ensure that most e-beam radiation is fully absorbed before reaching the sample (see see Suppl. fig. 7). This format would be compatible with the electron microscope vacuum chamber; as would biological samples placed in electron microscope-compatible liquid cells \cite{Liqcell1, Liqcell2, Liqcell3}. 


\clearpage

\clearpage

\section{Conclusion}

In conclusion, we have demonstrated a nanoscopic thermometry scheme based on the cathodoluminescence of NaYF$_4$: Yb$^{3+}$, Er$^{3+}$ nanoparticles, shown to be temperature-dependent at the sub-$\mu$m level. The mean cathodoluminescence ratio of two Er$^{3+}$ transitions could not, in our available data, be shown to yield reproducible changes contingent on the temperature of a sub-$\mu$m nanoparticle patch, presumably due to damage caused by prolonged e-beam irradiation. The lifetimes of such transitions, after fast e-beam blanking, were shown to depend on temperature in a linear fashion, yielding a measurement sensitivity on the order of $\sim$ 30 mK. The key advantage of cathodoluminescence-based lifetime thermometry stems from the fact that excited-state lifetimes are independent of variations in cathodoluminescence collection efficiency and background level, and of e-beam damage (all of which only alter signal-to-noise levels). In addition, and in contrast to intensity-based thermometry, there is no need to correlate multiple spectral lines. At the used e-beam energy of 10 keV, the cathodoluminescence edge spatial resolution is comparable to the electron signal's, even though the photon signal does not only stem from the directly excited region, but from a broader area $\sim$ 750 nm in radius; this figure is limited by cathodoluminescence excitation away from the e-beam scanning point due to electron scattering but is still superior to the resolution of excitation by a diffraction-limited optical Gaussian beam. It can moreover be improved by using lower-energy e-beams (for example,\ enabled by a higher-efficiency photon collection apparatus); it was calculated that a 2 keV e-beam would yield a resolution compared to the nanoparticles' size. We envision the implementation of cathodoluminescence-based thermometry with NaYF$_4$: Yb$^{3+}$, Er$^{3+}$ in two ways. First, using a single nanoparticle as a thermometer (on top of a sample that can withstand e-beam irradiation, such as miniaturized electronics) can immediately provide nanometric spatial resolution with the method described here. Second, a sheet of nanoparticles covering a sample, or a NaYF$_4$: Yb$^{3+}$, Er$^{3+}$ thin film under which a sample is deposited (or that is fabricated over a sample), can also function as a barrier against direct e-beam irradiation, thereby providing a platform for temperature diagnostics of delicate and biological specimens. The experiments reported here thus represent the first steps towards non-contact cathodoluminescence-based thermometry at the nanoscale.


\section{Sample preparation}

A sample of commercial NaYF$_4$: Yb$^{3+}$, Er$^{3+}$ (Mesolight Inc.) was drop cast on top of a silicon chip with $\sim$ 200 nm of silicon dioxide (deposited by plasma-enhanced chemical vapor). It was left to dry under ambient conditions.

Before the first electron microscopy session, the sample was gently cleaned for 2 minutes in an oxygen plasma in order to remove putative chemical residuals causing spurious cathodoluminescence emission (see Suppl. fig. 1). The same sample is used in all experimental runs, being subject to many thermal cycles.

\section{Cathodoluminescence measurements}

A Zeiss Gemini Supra 55 SEM is retrofit with a home-built parabolic mirror \cite{Kaz} in order to enable cathodoluminescence measurements. Custom software (ScopeFoundry \cite{scopefoundry, scopefoundry2}) synchronizes and controls a Raith 50 electrostatic beam blanker that can nominally pulse the e-beam with duty cycles as short as 1 $\mu$s; measured rise times are $\lesssim$ 25 ns. 

The specimen's temperature was changed using a home-built, open-loop heater stage placed inside the electron microscope's vacuum chamber. A metal clip pressed the chip containing the sample onto a copper surface; a K-type thermocouple was spot-welded to the inside of the metal clip, and as such was in contact with the sample side of the chip at all times. Thermal carbon paste was placed on the silicon chip, whose surface was in contact with the copper. Voltage readings from the thermocouple were amplified and automatically collected with the same frequency as the voltage readings yielded by the secondary electron detector. Voltage readings from the thermocouple were treated in the same fashion as the cathodoluminescence photon signal (eg.,\ frame averaging) before conversion to temperature.

A Hamamatsu R6094 (Hamamatsu H7421) PMT is used to record photons within the green (red) wavelength band. Bandpass filters are placed in the optical path before the PMTs (550/32 nm Semrock Brightline FF01-550/32-25 and 650/54 nm Semrock Brightline FF01-650/54-25). Absolute cathodoluminescence intensities vary on a daily basis given the apparatus' optical alignment and sensitivity. 

Both analog signals from the SEM's secondary electron detector, and digital signals from the PMTs are recorded by a National Instruments PXIe-6363 card.

Spectrally resolved cathodoluminescence measurements have been taken using a Ocean Optics QE65 spectrometer.

\section{Acknowledgments}

This work was supported by the Chemical Sciences, Geosciences and Biosciences Division, Office of Basic Energy Sciences, Office of Science, U.S. Department of Energy, FWP number SISGRN. Cathodoluminescence measurements at the Lawrence Berkeley National Laboratory Molecular Foundry were performed as part of the Molecular Foundry user program, supported by the Office of Science, Office of Basic Energy Sciences, of the U.S. Department of Energy under Contract No. DE-AC02-05CH11231. The preparation of this manuscript was supported by STROBE, a National Science Foundation Science $\&$ Technology Center under Grant No. DMR 1548924. C.\ D.\ Aiello acknowledges insightful discussions with Ayelet Teitelboim and Boerge Hemmerling, and assistance with the heater stage from Connor Bischak. A.\ D.\ Pickel and R.\ B.\ Wai acknowledge NSF Graduate Research Fellowships (DGE 1106400). N.\ S.\ Ginsberg acknowledges an Alfred P. Sloan Research Fellowship, a David and Lucile Packard Foundation Fellowship for Science and Engineering, and a Camille and Henry Dreyfus Teacher-Scholar Award.  

\clearpage


\providecommand{\latin}[1]{#1}
\providecommand*\mcitethebibliography{\thebibliography}
\csname @ifundefined\endcsname{endmcitethebibliography}
  {\let\endmcitethebibliography\endthebibliography}{}

\end{document}


\clearpage

\section{Characterization of experimental setup}

\begin{figure}[h!]
  \centering
  \includegraphics[width=1\textwidth]{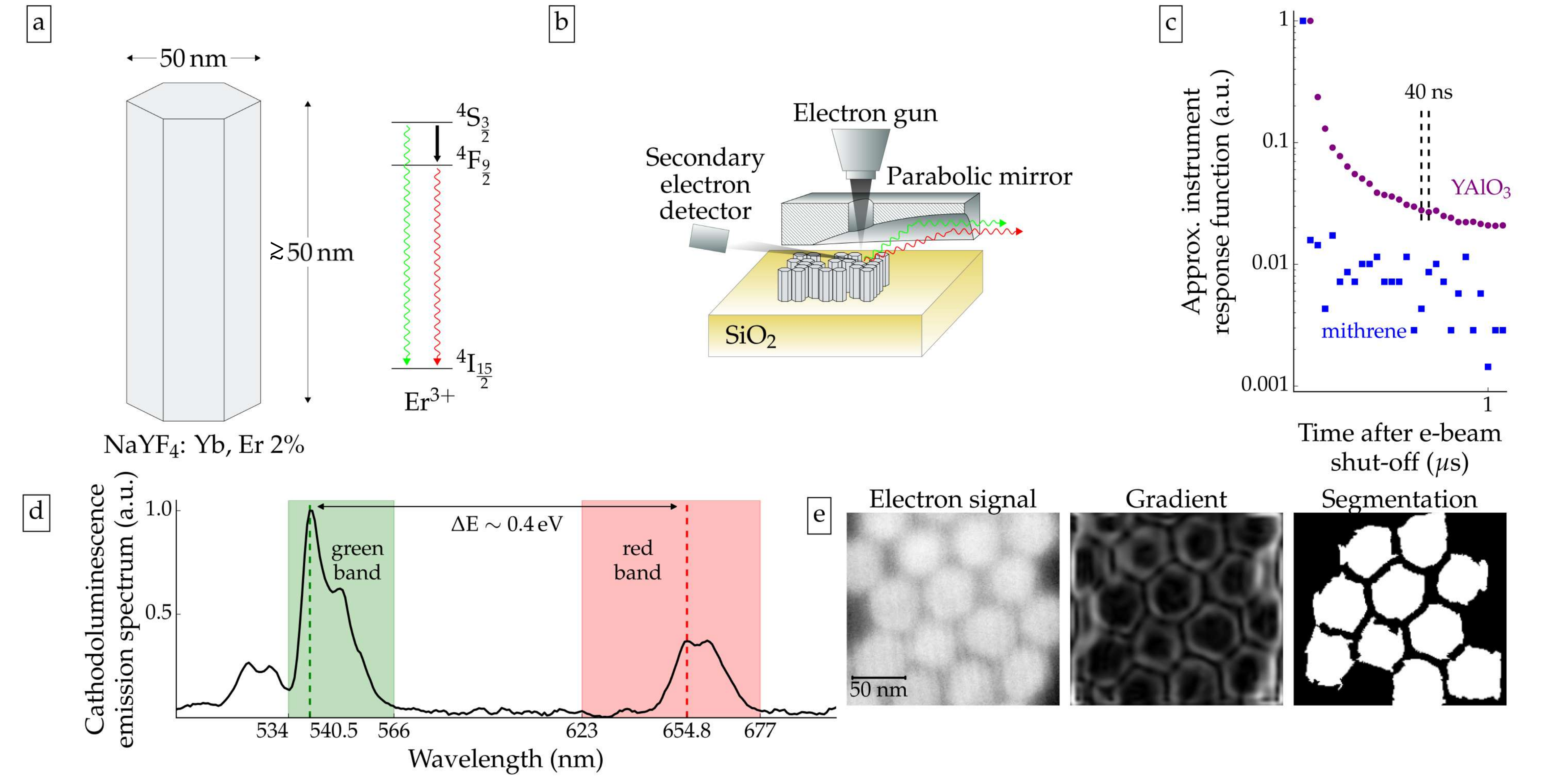}
  \caption{\textbf{The setup enables quantitative cathodoluminescence.} \mybox{\textbf{a}} The cathodoluminescence properties of lanthanide-doped hexagonal nanoparticles (nominally 50 nm in width, $\gtrsim$ 50 nm in length) are systematically probed. \mybox{\textbf{b}} The nanoparticles, prepared over a thin film ($\sim$ 200 nm) of SiO$_2$ on Si, are placed inside an SEM equipped with a parabolic mirror to collect cathodoluminescence. \mybox{\textbf{c}} The electron beam is electrostatically pulsed (rise time $\lesssim$ 25 ns) as it is scanned over the sample, and single-pixel synchronized photon counting can be achieved over time intervals down to 40 ns. After electron beam blanking, cathodoluminescence counts from a ps-lifetime metal-organic chalcogenide scintillator (mithrene \cite{mithrene}, blue squares) are a proxy for the instrument response function, which is approximately a delta function on the scale of our achievable 40 ns time intervals. We also track the cathodoluminescence decay of a ns-lifetime scintillator (YAlO$_3$ doped with Ce$^{+3}$, purple circles); for better display, this second decay curve is shifted to the right by 40 ns. \mybox{\textbf{d}} The measured cathodoluminescence spectrum of the lanthanide-doped nanoparticles is similar to the well-known photoluminescence spectrum (eg.,\ references \cite{solarcell, TM1, theo2, TM3}), confirming that the emission is mainly from the Er$^{3+}$ emitter. Here, we study cathodoluminescence in two optical bands separated by $\sim$ 0.4 eV (red and green shaded areas, boundaries determined by the bandpass filters); they represent known optical decays in the Er$^{3+}$ ions ($^4$F$_{\frac{9}{2}}$ $\rightarrow$ $^4$I$_{\frac{15}{2}}$ and $^4$S$_{\frac{3}{2}}$ $\rightarrow$ $^4$I$_{\frac{15}{2}}$ respectively, see \mybox{\textbf{a}}). \mybox{\textbf{e}} Cathodoluminescence excitation is not restricted to the spot irradiated by the primary e-beam, due to the lateral spread of back-scattered electrons returning to the surface from deeper inside the substrate; nevertheless, cathodoluminescence information as centered at individual nanoparticles can be monitored at the single-pixel level by using the gradient of the synchronously acquired electron signal as a mask for segmentation.}
  \label{Fig1}
\end{figure}

We study the cathodoluminescence from NaYF$_4$ nanoparticles co-doped with Yb$^{3+}$ and Er$^{3+}$, in particular the Er$^{3+}$ energetic transitions $^4$S$_{\frac{3}{2}}$ $\rightarrow$ $^4$I$_{\frac{15}{2}}$ and $^4$F$_{\frac{9}{2}}$ $\rightarrow$ $^4$I$_{\frac{15}{2}}$ (green and red, respectively; measured at a low magnification of 1k X in Suppl. fig.\ \ref{Fig1}\mybox{\textbf{d}}). The nanoparticles have the shape of hexagonal rods, nominally of 50 nm diameter, $\gtrsim$ 50 nm height, as depicted in Suppl. fig.\ \ref{Fig1}\mybox{\textbf{a}}.

As seen in Suppl. fig.\ \ref{Fig1}\mybox{\textbf{b}}, nanoparticles drop cast on a silicon chip with a thin silicon oxide layer are excited by a field-emission scanning electron microscope (SEM) equipped with a custom-built in-vacuum mirror.

An approximation of the instrument response function is obtained by recording the transient cathodoluminescence of phosphors decaying much faster than the minimum photon-collection integration time in the setup (40 ns, limited by the data acquisition card). The cathodoluminescence of the metal-organic chalcogenide `mithrene' (silver benzeneselenolate) \cite{mithrene0, mithrene}, expected to have $\sim$ ps luminescence lifetimes, decays within 40 ns by almost two orders of magnitude after the e-beam is shut off, constituting a good approximation of a delta function at the ns scale, as depicted in Suppl. fig.\ \ref{Fig1}\mybox{\textbf{c}} (blue squares). Similarly, we can track the exponential cathodoluminescence decay of known standard phosphors with $\sim$ ns lifetimes such as YAlO$_3$ doped with Ce$^{+3}$ (purple circles) \cite{coconnor}. 

The secondary electron signal is segmented using a gradient-sensitive algorithm \cite{canny}; the process is depicted in Suppl. fig.\ \ref{Fig1}\mybox{\textbf{e}}. Cathodoluminescence is excited throughout a region whose size depends on the extent of the electron scattering (both incident and backscattered), which in turn depends on the used e-beam energy; as such, segmentation of the cathodoluminescence signal only yields a weighted-averaged
information of the cathodoluminescence at such excitation regions; see discussion around Fig.\ 4 in the main text.

\clearpage

\clearpage

\section{Spectrally-resolved cathodoluminescence \newline of NaYF$_4$: Yb$^{3+}$, Er$^{3+}$}

\begin{figure}[h!]
  \centering
  \includegraphics[width=1\textwidth]{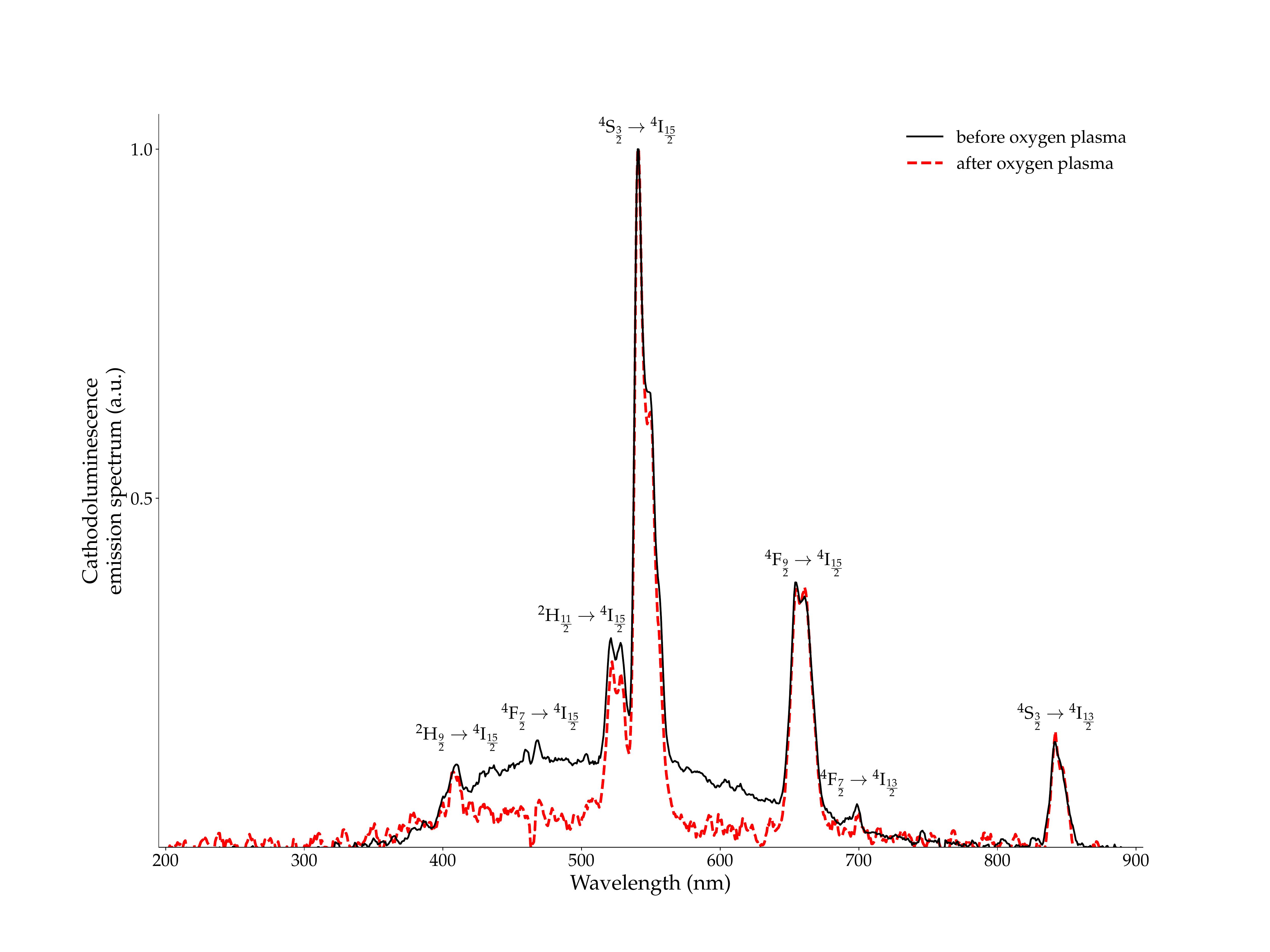}
  \caption{\textbf{The cathodoluminescence spectrum of NaYF$_4$: Yb$^{3+}$, Er$^{3+}$ is dominated by Er emission.} The full cathodoluminescence spectrum of the nanoparticles was measured at room-temperature (dashed red line) with a spectrometer at 1k X magnification (patch of size 0.01 mm$^2$). Peaks correspond to Er emission lines (transitions assigned). Before a 2-minute treatment with oxygen plasma, the spectrum presents a broad background (solid black line) which we hypothesize to be caused by the presence of cathodoluminescent chemical residuals.}
  \label{Figa}
\end{figure}

\clearpage

\section{Quantitative cathodoluminescence intensities and lifetimes}

We first investigate the cathodoluminescence of single-layer patches of nanoparticles as a function of increasingly longer e-beam excitations. Unless explicitly stated otherwise, standard imaging parameters are: 10 keV e-beam energy, 379 pA beam current through a 30 $\mu$m aperture, pixel step of 2.48 nm (corresponding to a magnification of 150k X), photon integration time of 1 $\mu$s, and transient cathodoluminescence collection after e-beam blanking of at least 500 $\mu$s.

\begin{figure}[h!]
  \centering
  \includegraphics[width=1\textwidth]{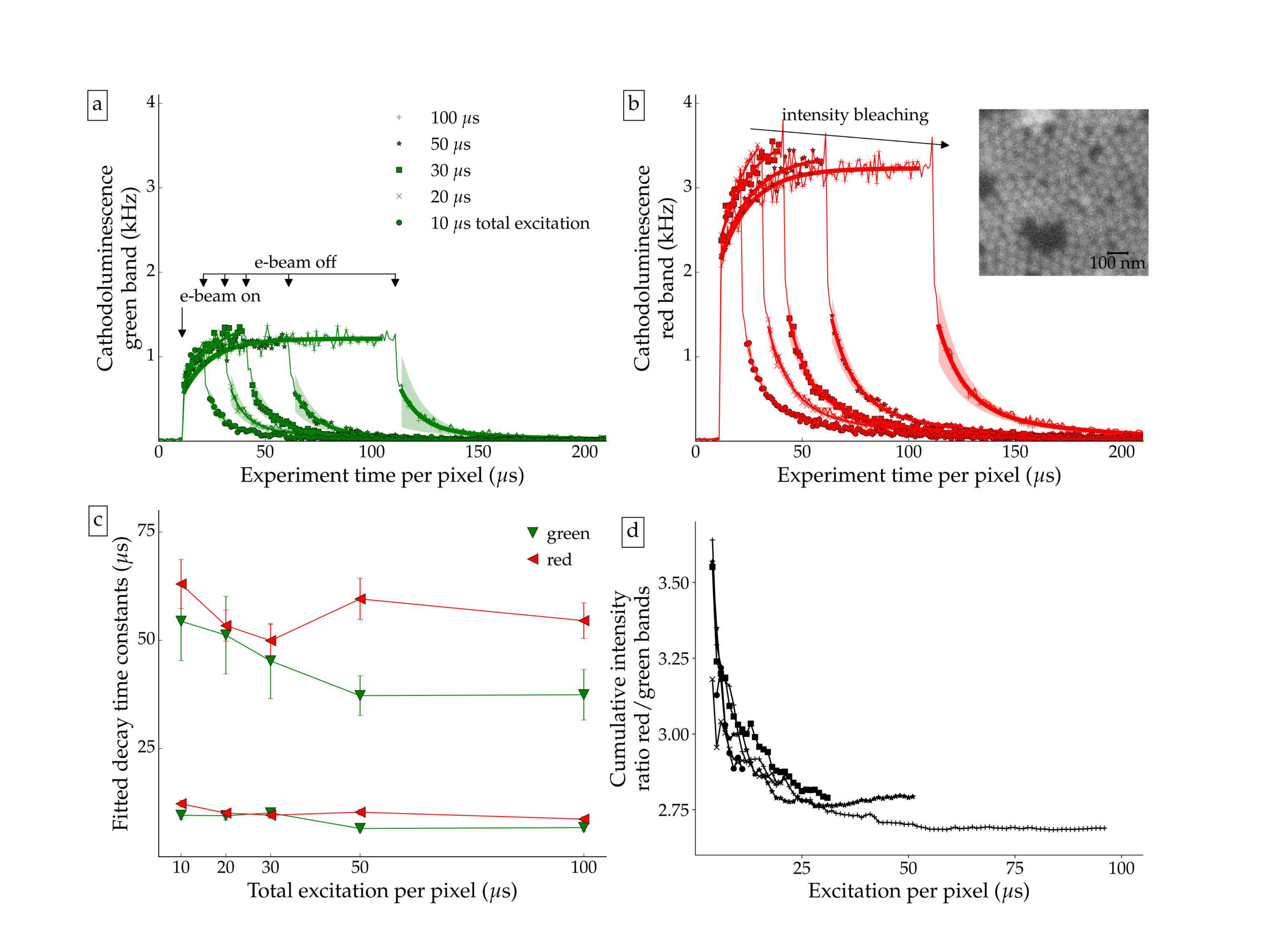} 
  \caption{\textbf{Cathodoluminescence intensities and lifetimes can be monitored.} \mybox{\textbf{a}}, \mybox{\textbf{b}} Cathodoluminescence of a sub-$\mu$m patch of nanoparticles (inset), for increasing total e-beam excitation durations, can be fitted for the rise time (single exponential) and excited-state lifetimes (double exponential). These models were chosen to minimize fitting errors. Whereas different excitation times do not considerably change the cathodoluminescence decay dynamics, the mean cathodoluminescence intensity emission decreases for both light bands (indicated by the arrow in \mybox{\textbf{b}}) as a result of electron beam bleaching. \mybox{\textbf{c}} We study the relaxation process following fast electron beam blanking. Fitted decay time constants for the different curves are within two standard deviations of each other (vertical extent of shown errorbars), except for one point (red band, 50 $\mu$s excitation, longer time constant). \mybox{\textbf{d}} Even with bleaching, the intensity ratio between light bands (while the electron beam is on) follows the same decreasing trend for increasing excitation durations, making this quantity a stable metric; the ratio is observed to stabilize at $\sim$ 2.75 for excitations longer than typical cathodoluminescence rise times ($\gtrsim$ 30 $\mu$s); this value is determined by the relative alignment of the PMTs, and can be calibrated. Here, the curves for the different electron beam excitation durations are plotted using the same markers as in \mybox{\textbf{a}}, \mybox{\textbf{b}}.}
  \label{Fig2}
\end{figure}

A $\sim$ 0.53 $\mu$m$^2$ sample region (inset of Suppl. fig.\ \ref{Fig2}\mybox{\textbf{b}}) is consecutively excited by the e-beam for increasing total durations of $\{$10, 20, 30, 50, 100$\}$ $\mu$s per pixel. Cathodoluminescence is recorded for each pixel: for 10 $\mu$s prior to the fast electrostatic switch of the e-beam; through the e-beam excitation; and after the beam is switched off. The topographic image generated by the secondary electron signal is then segmented in regions with (bright) and without (dark) nanoparticles using a two-Gaussian mixture model \cite{gmm}; bright pixels are averaged together, yielding the cathodoluminescence signal as a function of experiment time per pixel, which is depicted in Suppl. fig.\ \ref{Fig2}\mybox{\textbf{a}} and \mybox{\textbf{b}}, for the green and red light bands. 

Rise and decay transients as the e-beam is turned on and off are not expected to change as a function of total electron dose. Cathodoluminescence rise and decay behaviors are fitted, respectively, to single and double exponentials plotted as solid lines in Suppl. fig.\ \ref{Fig2}\mybox{\textbf{a}} and \mybox{\textbf{b}}. Such fitting functions were empirically chosen for having the highest F-statistic. The shaded areas around the decay curves represent the error in the model at one standard deviation (see below for definition of error in the model). Fitted decay time constants are shown in Suppl. fig.\ \ref{Fig2}\mybox{\textbf{c}}; except for one point (red band, 50 $\mu$s excitation, longer time constant), they all fall within two standard deviations of each other (as indicated by the errorbars, although some are smaller than the marker's size). Longer decay time constants are on the same order of magnitude as previously reported under photoluminescence \cite{TM1, theo2}. We do not believe that shorter decay time constants are beam-blanking related, as beam blanking occurs much faster (ns) than the photon integration time ($\mu$s). Fitted rising time constants have been omitted, as they reflect a driven process occurring in the presence of the e-beam; we chose to concentrate on the purely relaxation process related to the excited-state lifetimes instead.

In contrast to the near dose-independence of the lifetimes (Suppl. fig.\ \ref{Fig2}\mybox{\textbf{c}}, largely within error bars), the mean emitted intensity under continuous excitation depends on electron dose and is observed to bleach in both light bands (see arrow in Suppl. fig.\ \ref{Fig2}\mybox{\textbf{b}}). Despite bleaching of the sample due to electron dose, we establish a metric using intensity ratios of different emission wavelength ranges. This metric is stable even under bleaching. Such a metric is useful if it can be shown to change only under specific experimental conditions; for example, optically-excited thermometry using NaYF$_4$: Yb$^{3+}$, Er$^{3+}$ relies precisely on the fact that the ratio of spectral peak intensities changes as a function of the nanoparticle temperature \cite{TM1, TM2, TM3, TM4}. We plot the cumulative cathodoluminescence intensity ratio for increasing excitation durations in Suppl. fig.\ \ref{Fig2}\mybox{\textbf{d}}; this quantity follows the same tendency for each experimental run in a series of several; it stabilizes at $\sim$ 2.75 for the considered experimental conditions, and for irradiation longer than typical cathodoluminescence rise times ($\gtrsim$ 30 $\mu$s, similar to reported under photoluminescence \cite{TM1}). Thus, assuming that the effect of intensity bleaching is cumulative, the cathodoluminescence intensity ratio can be used as a metric at least for total irradiation durations up to 210 $\mu$s per pixel (i.e.\ taking into account the fact that, after these experiments, each pixel within the considered region had been irradiated for 10 $+$ 20 $+$ 30 $+$ 50 $+$ 100 $=$ 210 $\mu$s).

We note that, in these plots, the mean cathodoluminescence intensity is higher for the red band, in contrast with the spectrally resolved signal of \ Suppl. fig.\ \ref{Fig1}\mybox{\textbf{d}}. This is due to different sensitivities and optical alignments of the PMTs for each color channel, which vary from day to day in our experiments. The PMTs, however, are significantly more sensitive than the available spectrometer, which enabled experiments with higher spatial resolution. With a simple calibration, a stable cathodoluminescence intensity ratio can be defined regardless of the particular relative alignment of the PMTs.

\clearpage

\clearpage

\section{Cathodoluminescence intensity dependence \\ on standard electron microscopy parameters}

It is relevant to characterize the cathodoluminescence intensity of the nanoparticles as a function of electron microscopy parameters \cite{ogletree, TM2}. To do so, comparable patches of nanoparticles (size 0.56 $\mu$m$^2$, containing on average 137 $\pm$ 23 nanoparticles; this corresponds to a nominal nanoparticle filling fraction $\sim$ 40.2 $\pm$ 6.8$\%$) not previously exposed to the e-beam were excited for a duration of 500 $\mu$s per pixel, 
for different pixel steps, e-beam currents and energies, using the same microscope aperture of 30 $\mu$m. In each scan, only one parameter was varied from its standard value (namely, 10 keV e-beam energy, 379 pA beam current, pixel step of 2.48 nm). 

The cathodoluminescence intensity ratio established in Suppl. fig.\ \ref{Fig2}\mybox{\textbf{d}} is then obtained by integrating the photon signal per pixel while the e-beam is on, and by subsequent averaging over all pixels. This ratio is then normalized to the benchmark ratio obtained using the standard set of parameters; this benchmark ratio (i.e. the denominator used in the normalizations) varies from day to day given different PMT relative alignments, but it has been observed in our setup to lie between $\sim$ 2 and 3.5 for excitation durations longer than $\sim$ 30 $\mu$s per pixel; for example, in the experiments of Suppl. fig.\ \ref{Fig2}, the benchmark ratio was found to be $\sim$ 2.75, as seen in the long-time asymptote of Suppl. fig.\ \ref{Fig2}\mybox{\textbf{d}}. 

\begin{figure}[h!]
  \centering
  \includegraphics[width=1\textwidth]{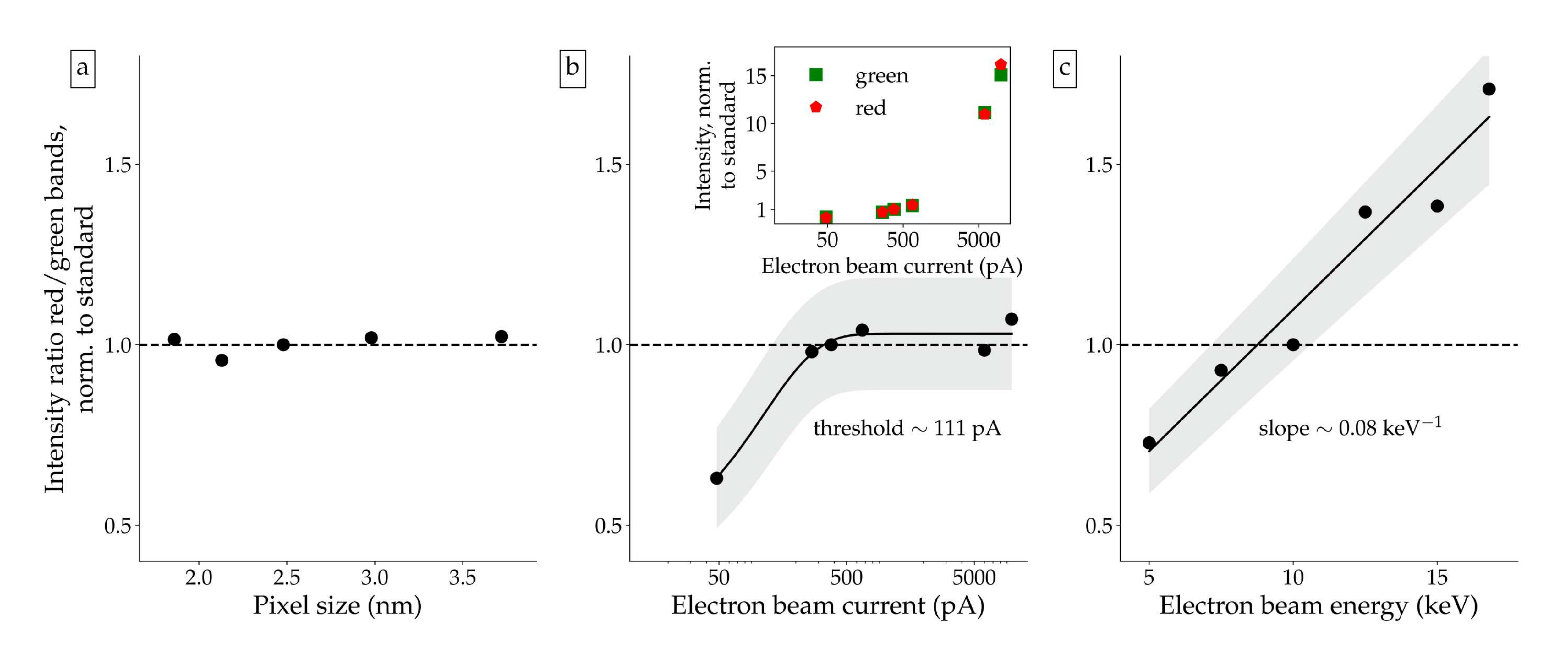}
  \caption{\textbf{The cathodoluminescence intensity is characterized as a function of electron microscopy parameters.} All data points are normalized to the baseline case which uses the following parameters: 2.48 nm pixel, 379 pA beam current, and 10 keV beam energy. Since the individual data points are the result of averaging over $\sim$ 10$^5$ pixels, the standard deviation of the plotted data points are much smaller than the marker size and are thus omitted. Where present, gray areas are the error in the model at one standard deviation (see Supplementary information for definition of error in the model). \mybox{\textbf{a}} The intensity ratio is independent of pixel size (a proxy for electron dose per nanoparticle). \mybox{\textbf{b}} Above a threshold value presumably related to signals too small to be reliably recorded, varying the electron beam current does not alter the nanoparticle cathodoluminescence emission ratio even as raw intensities differ by more than one order of magnitude (inset). This behavior is expected. The exponential function fitted to the data yields a threshold current (fitted exponent) of $\sim$ 111 pA. \mybox{\textbf{c}} The intensity ratio depends on the electron beam energy from 5 to 20 keV in an approximately linear fashion, with slope $\sim$ 0.08 keV$^{-1}$; we speculate that higher voltages can excite more phonon modes, and accordingly modify the emission of the transition more strongly coupled to such modes -- in this case, the red light band. This would increase the intensity ratio as observed.}
  \label{Fig3}
\end{figure}

The normalized cathodoluminescence ratio is shown not to depend on pixel size (i.e.\ on the electron dose per nanoparticle) over the recorded range, covering magnifications from 100 to 200k X; this is depicted in Suppl. fig.\ \ref{Fig3}\mybox{\textbf{a}}. Beyond an exponential best-fit beam current threshold of $\sim$ 111 pA (which is dependent on the apparatus' photon collection efficiency), the intensity ratio is approximately constant over more than an order of magnitude increase in current (Suppl. fig.\ \ref{Fig3}\mybox{\textbf{b}}) -- even as the raw band intensities increase by a factor of up to 15 (inset). This is due to the fact that, beyond a certain low-level signal threshold that is not properly detected by the PMTs, increasing or decreasing the absolute number of electrons exciting the sample per time interval does not change the relative cathodoluminescence yields -- thus keeping the intensity ratio constant. This finding is especially relevant since beam currents tend to fluctuate in time with filament usage, and further supports the claim that the cathodoluminescence intensity ratio is a stable metric under a variety of experimental conditions. However, changes in e-beam energy are measured to influence the intensity ratio in a linear fashion for the measured range from 5 to 20 keV (Suppl. fig.\ \ref{Fig3}\mybox{\textbf{c}}), with a fitted slope of $\sim$ 0.08 keV$^{-1}$. We posit that this phenomenon might be related to the fact that an increased e-beam energy can populate more phonon modes, and thus influence more strongly the emission of the transition more strongly coupled to such modes -- in this case, the red light band \cite{Chan12,theo}, thereby increasing the intensity ratio.

\clearpage

\clearpage

\section{Thermometry mediated by cathodoluminescence intensity: \newline spectrally resolved at low magnification (1k X) }

\begin{figure}[h!]
  \centering
  \includegraphics[width=1\textwidth]{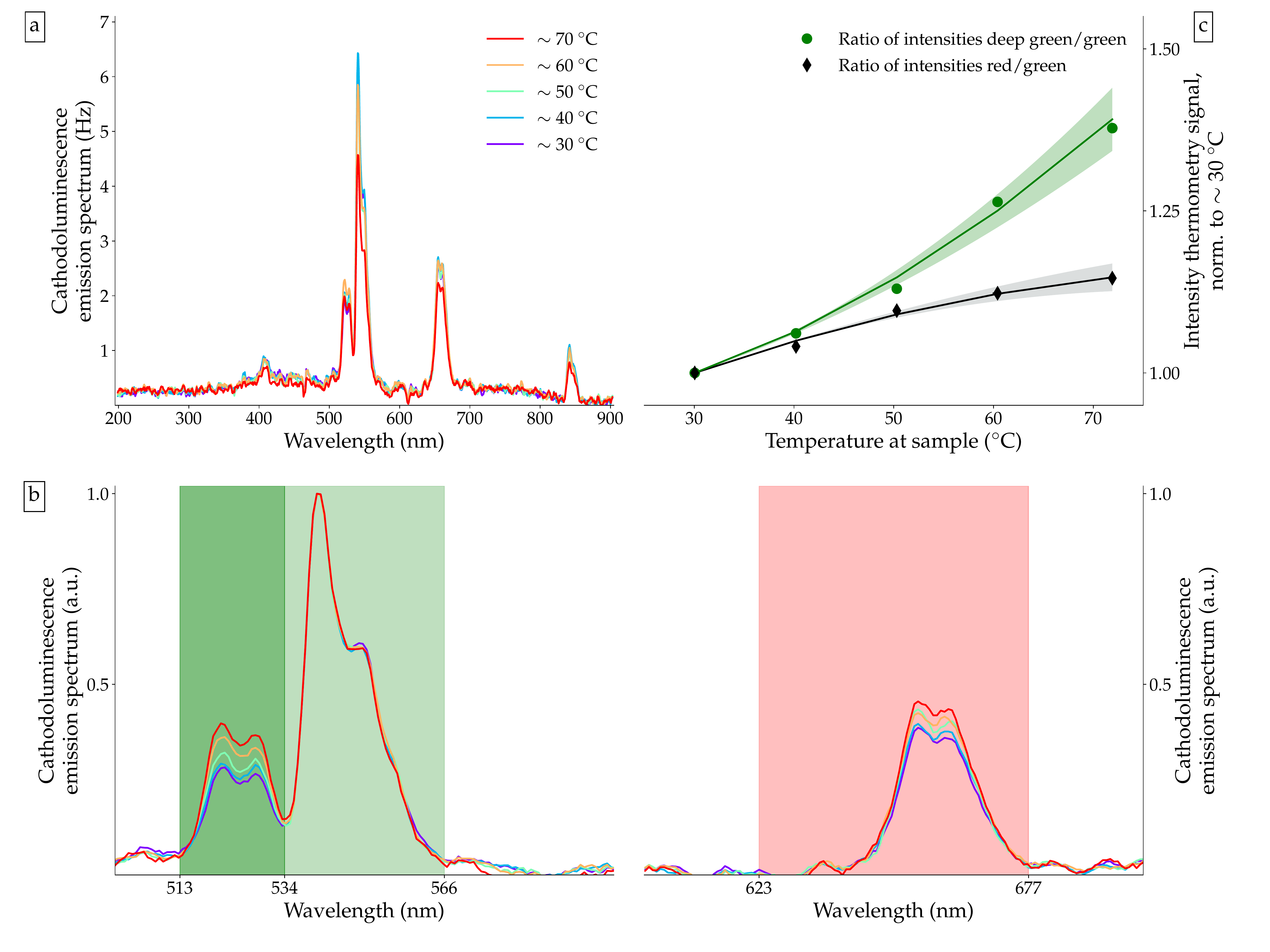}
  \caption{\textbf{Quantitative cathodoluminescence can find an application in thermometry.} \mybox{\textbf{a}} The spectrally resolved cathodoluminescence emission of the NaYF$_4$: Yb$^{3+}$, Er$^{3+}$ nanoparticles as taken at 1k X magnification is shown to depend on the temperature of the sample. \mybox{\textbf{b}} Zoomed cathodoluminescence emission as a function of temperature at wavelengths of interest. These are: the red and green regions corresponding to the filters used in the experimental data recorded with PMTs (shaded red and green, respectively); and an additional deep green region (513--534 nm) often used in the context of optical thermometry with such phosphors. Here, for easier visualization, the values are normalized, with a value of 1 corresponding to the peak of the green transition at all temperatures. The scale of the wavelength axis is the same for both plots. \mybox{\textbf{c}} Using the spectrometer readings at different temperatures, a thermometry signal can be defined by taking the integrated intensity ratios at the different spectral regions, recorded at increasing temperatures. This signal can change up to $\sim$ 40$\%$ in a 40 $^{\circ}$C span, with sensitivities $\sim$ 0.10--1.48$\%$ $^{\circ}$C$^{-1}$, in good agreement with those reported in optical measurements on similar nanoparticles \cite{TM1, TM2, Diaz, kilbane}. Shading represents the error in the model at the level of one standard deviation.}
  \label{Fig5}
\end{figure}

The quantitative cathodoluminescence study of nanophosphors can be applied to thermometry at the nanoscale. In Suppl. fig.\ \ref{Fig5}\mybox{\textbf{a}} and \mybox{\textbf{b}}, the spectrally resolved cathodoluminescence is recorded for 60 s with the e-beam continuously on, at 1k X magnification (patch of size 0.01 mm$^2$), at each of 5 temperatures. The beam is otherwise blanked. The sample's temperature is controllably increased from $\sim$ 30 $^{\circ}$C up to $\sim$ 70 $^{\circ}$C at a rate of $\sim$ 10 $^{\circ}$C per hour. In terms of dose, 5 min of cumulative imaging at 1k X magnification (estimated dose $\sim$ 500 pC/$\mu$m$^2$) is in bounds with the dose span previously covered by the higher-current data points of Suppl. fig.\ \ref{Fig3}\mybox{\textbf{b}} (500 $\mu$s of exposure at 150k X magnification).

Even if the scattering rate of the $^4$F$_{\frac{9}{2}}$ $\rightarrow$ $^4$I$_{\frac{15}{2}}$ transition is affected by temperature changes, both the $^4$S$_{\frac{3}{2}}$ $\rightarrow$ $^4$I$_{\frac{15}{2}}$ and $^2$H$_{\frac{11}{2}}$ $\rightarrow$ $^4$I$_{\frac{15}{2}}$ transitions have significantly larger scattering rates at higher temperatures. This latter transition (513--534 nm, deep green) is known to be thermally coupled to the $^4$F$_{\frac{9}{2}}$ $\rightarrow$ $^4$I$_{\frac{15}{2}}$ transition \cite{Chan12}, so that a simple Boltzmann distribution governs the relative populations of the green and deep green transitions. 

We quantify the intensity thermometry signals from the ratios of the counts from different wavelength bands in Suppl. fig.\ \ref{Fig5}\mybox{\textbf{c}}. After normalizing these signals by the intensity ratios at $\sim$ 30 $^{\circ}$C, we empirically fit parabolas ($a$ $\cdot$ $T^2$ $+$ $b$ $\cdot$ $T$ $+$ $c$, higher F-statistic than a linear model) with a fixed point to them (the parabola is constrained to go through the data point at $\sim$ 30 $^{\circ}$C). For the red/green bands intensity ratios, we find $\{ a$ $=$ -4.33 $\pm$ 1.22 $\times$ 10$^{-5}$ $^{\circ}$C$^{-2}$ , $b$ $=$ 5.33 $\pm$ 0.44 $\times$ 10$^{-3}$ $^{\circ}$C$^{-1}\}$. Optical experiments, however, often exploit the luminescence intensity ratio of the coupled transitions in the green and deep green \cite{TM1}; the thermometry signal fits for these transitions yield $\{ a$ $=$ 9.59 $\pm$ 2.78 $\times$ 10$^{-5}$ $^{\circ}$C$^{-2}$, $b$ $=$ 5.17 $\pm$ 0.99 $\times$ 10$^{-3}$ $^{\circ}$C$^{-1}\}$. The shaded areas correspond to the error in the model to the level of one standard deviation. Importantly, both considered signals change by tens of percent within a 40 $^{\circ}$C span, corroborating the sensitivity of the method; this is quantitatively seen by taking the temperature derivative of the fit to the cathodoluminescence signal $S$, 
\begin{equation}
\frac{\partial S}{\partial T} = 2a \cdot T + b \ ,
\end{equation}
which yields a measure of changes in the signal with respect to temperature. In the 30--70 $^{\circ}$C range considered, this sensitivity has mean values of 0.10 and 1.48$\%$ $^{\circ}$C$^{-1}$ for the red/green and green/deep green transitions, respectively, which places the experimental results at the lower and upper bounds of reported sensitivities in related photoluminescence measurements ($\sim$ 0.10--1.50$\%$ \cite{TM1, TM2, Diaz, kilbane}).

\clearpage

\clearpage

\section{Definition of error in the model}

The error in the model represents one standard deviation of all fitting parameters. This approach assesses the fitting as a whole, not only as a function of a single parameter's statistics.

Given a model $f(\{a, b, c, d, ...\})$ where $\{a, b, c, d, ...\}$ are fitting parameters, and least-squares best fits are $\{a^\star, b^\star, c^\star, d^\star, ...\}$; the error in the model $f_e(\{a^\star, b^\star, c^\star, d^\star, ...\})$ is defined as the standard deviation of the model function calculated at all combinations $\{a^\star \pm \sigma_a$, $b^\star \pm \sigma_b, ...\}$, where $\sigma_i$ is the fitted parameter $i^\star$'s standard deviation.

In Fig.\ 2 in the main manuscript; in Suppl. figs.\ \ref{Fig2}\mybox{\textbf{a}} and \mybox{\textbf{b}}; \ref{Fig3}\mybox{\textbf{b}} and \mybox{\textbf{c}}; and \ref{Fig5}\mybox{\textbf{c}}, the error in the model has been plotted as shaded areas corresponding to $f(\{a^\star, b^\star, c^\star, d^\star, ...\}) \pm f_e(\{a^\star, b^\star, c^\star, d^\star, ...\})$.

\clearpage

\section{Calculation of lifetime thermometry sensitivity, $\partial T$}

Referring to the measurements of Fig.\ 3\mybox{\textbf{a}}, we apply linear fits to the temperature-dependent data $\tau_{\textrm{d}}(t_{\textrm{fix}},T)$ for each $t_{\textrm{fix}}$ in [0, 1405] $\mu$s, in 1 $\mu$s intervals. Counts obey Poisson statistics, so that the minimizing functional is given by
\begin{equation}
\textrm{min}\left(2 \sum \left(\tau_{\textrm{d}} \cdot ln(\tau_{\textrm{d}}) - \tau_{\textrm{d}}\cdot ln(\tau_{\textrm{m}}) - (\tau_{\textrm{d}} - \tau_{\textrm{m}})\right) \right) \ .
\end{equation}
Here, $\tau_{\textrm{m}}$ is the model (i.e.,\ $\tau_{\textrm{m}}(t_{\textrm{fix}}, T) = \alpha(t_{\textrm{fix}}) \cdot T + \beta(t_{\textrm{fix}})$), and $\tau_{\textrm{d}}$ are the calculated values for the characteristic lifetime obtained by using the experimental cumulative intensity counts data,
\begin{equation}
\tau_{\textrm{d}}(t_{\textrm{fix}}, T) \equiv \ddfrac{\int_{0}^{t_{\textrm{fix}}} I(t',T)t'dt'}{\int_{0}^{t_{\textrm{fix}}} I(t',T)dt'} \ .
\end{equation}
This is equivalent to minimizing the Poisson deviance \cite{princeton}. This procedure yields a series of $\alpha^{\star}(t_{\textrm{fix}})$, where the $^{\star}$ denotes a best-fit value.

For each $t_{\textrm{fix}}$, we can calculate the standard deviation $\sigma_{\tau}$ of the mean $\tau_{\textrm{d}}(t_{\textrm{fix}}, T)$ for all 6 temperatures $T$. 

Finally, the lifetime thermometry sensitivity is numerically obtained, for each $t_{\textrm{fix}}$, as
\begin{equation}
\partial T(t_{\textrm{fix}}) = \ddfrac{\sigma_{\tau_{\textrm{d}}(t_{\textrm{fix}}, \textrm{all } T)}}{|\alpha^{\star}(t_{\textrm{fix}})|} \ .
\end{equation}

\clearpage

\section{Lifetime thermometry results \newline for temperature-increasing data}

\begin{figure}[h!]
  \centering
  \includegraphics[width=1\textwidth]{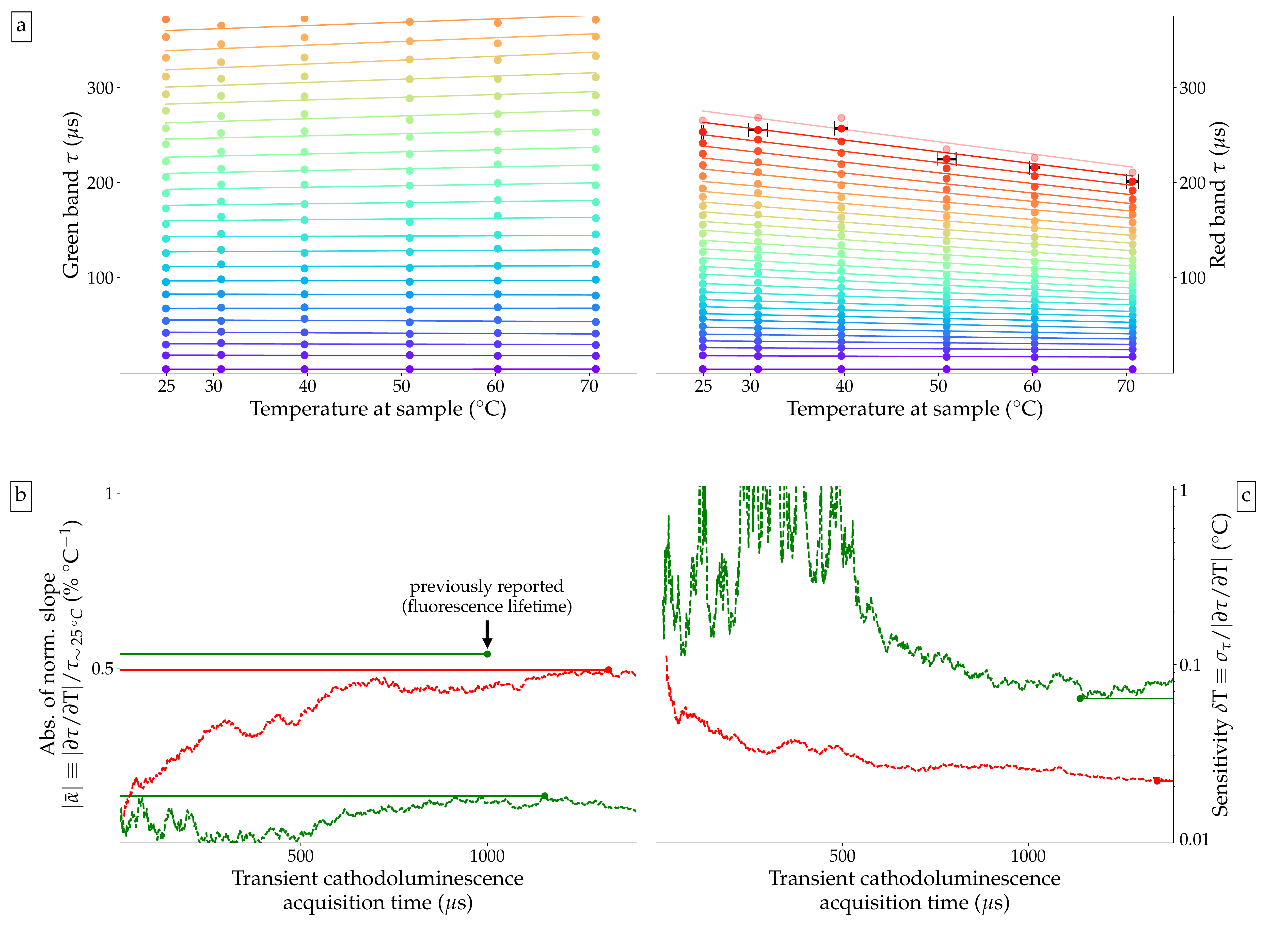}
  \caption{\textbf{Lifetime thermometry results are on the same order of magnitude for data taken with temperature increasing and decreasing.} This plot depicts data taken under identical conditions to Fig.\ 3 of the main text, but for temperatures that are increasing. Lifetimes, sensitivities and optimal acquisition times are on the same order of magnitude as those shown in Fig.\ 3 of the main text.}
  \label{nnn}
\end{figure}

\clearpage

\section{Monte-Carlo simulations of electron scattering}

\begin{figure}[h!]
  \centering
  \includegraphics[width=1\textwidth]{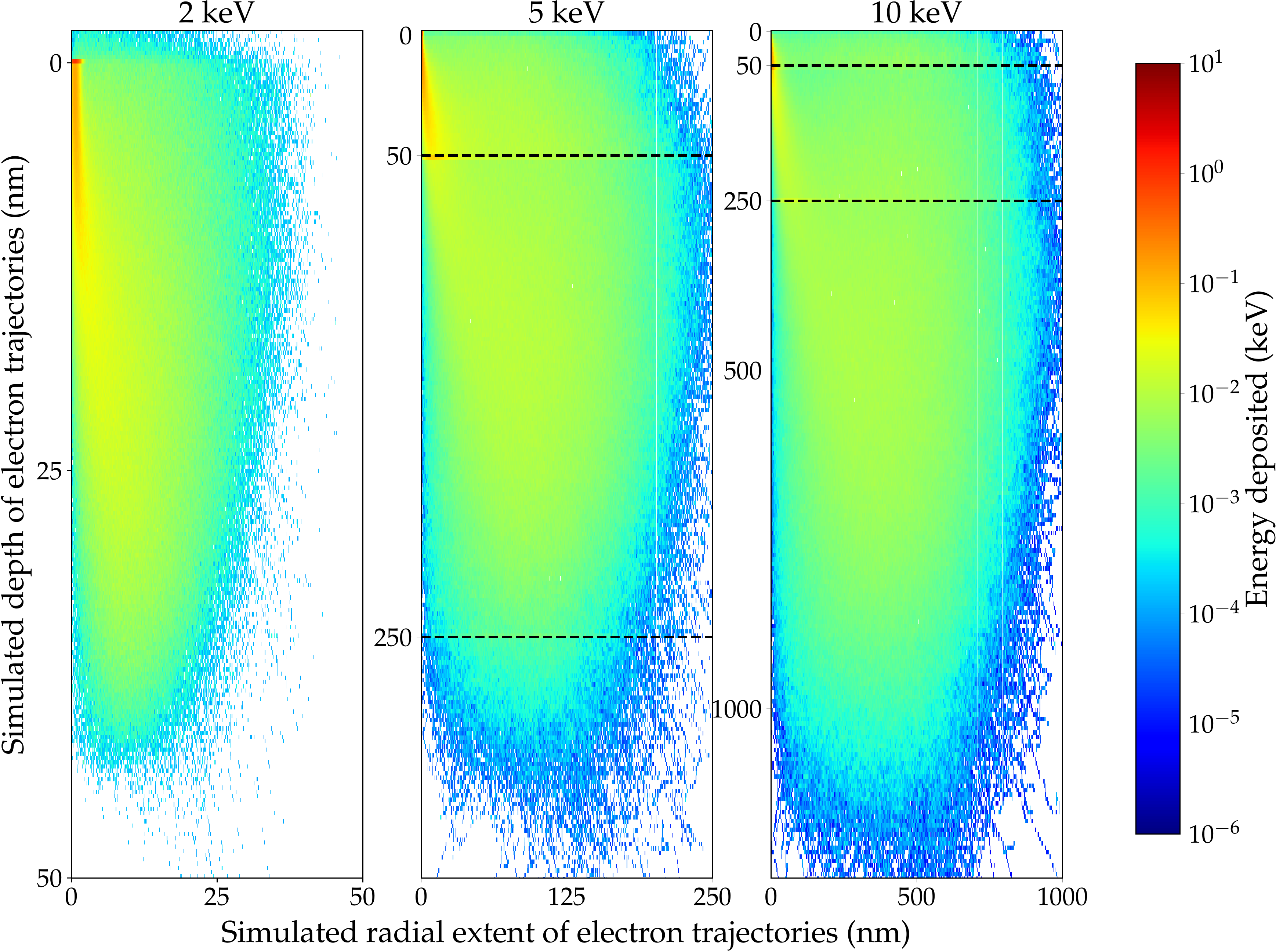}
  \caption{\textbf{Simulated deposited energy per depth and radius.} Depths of 50 and 250 nm are marked by dashed lines, representing the simulated interfaces: nanoparticle--oxide and oxide--silicon substrate.}
  \label{Fig11}
\end{figure}

Electron trajectories and energy loss have been simulated using a numerical Monte-Carlo algorithm using the free Casino software \cite{CASINO}. A layered material composed of NaYF$_4$ (50 nm in depth, density 4.20 g/cm$^3$), SiO$_2$ (200 nm in depth, density 2.65 g/cm$^3$) and Si (substrate, density 2.33 g/cm$^3$) was used to determine the depths at which electron beams of different accelerating voltages deposit energy. The electron beam was modeled to be 2 nm in diameter. 10$^5$ trajectories in a beam centered at a depth of 0 and at a radius of 0 were recorded for individual electrons; scattering throughout the material is followed until the electrons' final energies reach at most 1 eV (or $\sim$ 1.2 $\mu$m in wavelength, which is lower in energy than the synthetizer Yb absorption at $\sim$ 980 nm). The lateral extent of electron scattering was not restricted in any way in the simulations. The average energy deposited (energy lost in between scattering events) is calculated at spatial bins for certain depths and radii. The maximal bin values were chosen as the extrema of radius and depth yielded by the simulations. For all e-beam energies, there were 250 depth and 500 radial bins. This map of deposited energies is shown in Suppl.\ fig.\ \ref{Fig11}.

The cathodoluminescence point spread functions as depicted in Fig.\ 4\mybox{\textbf{c}} of the main text are calculated by assuming that, at every simulation step, each electron with energy higher than 1 eV can generate one cathodoluminescence excitation at any depth within the first 50 nm in the layered material, corresponding to the nanoparticles' nominal height. Essentially, this assumption takes the interaction cross-section for an electron to initiate a cathodoluminescence process to be independent of the energy of the electron; a similar calculation assuming that cathodoluminescence intensity is instead proportional to the energy deposited at different depths yields qualitatively similar results, with minimally improved point spread functions. In Fig.\ 4\mybox{\textbf{c}}, we chose to depict the results using the most stringent calculation.

\clearpage

\clearpage

\section{Simulated cathodoluminescence profile \newline of an array of nanoparticles}

\begin{figure}[h!]
  \centering
  \includegraphics[width=1\textwidth]{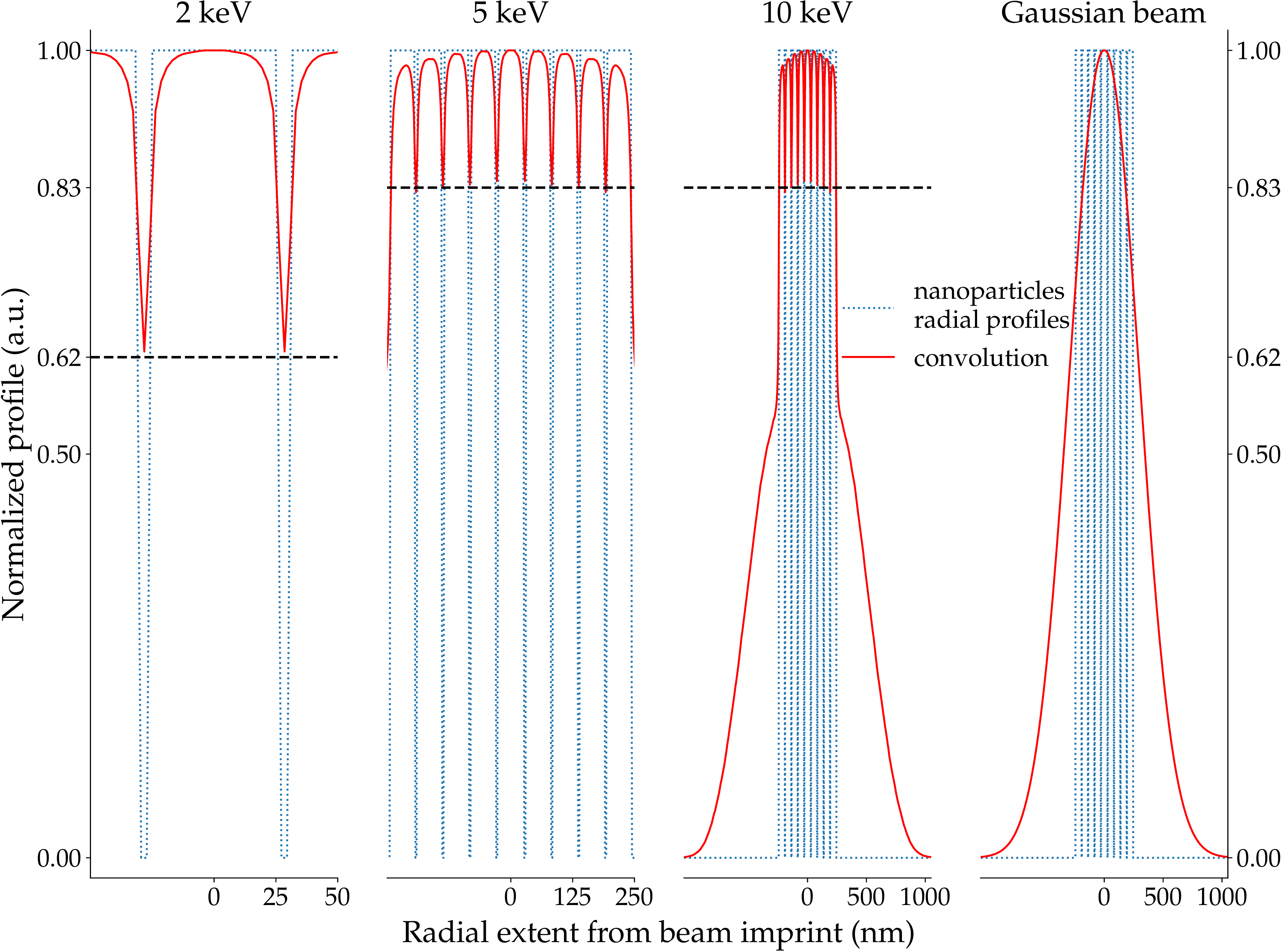}
  \caption{\textbf{Simulated cathodoluminescence profile of an array of nanoparticles.} These are calculated by convolving the point spread function of Fig.\ 4\mybox{\textbf{c}} with the radial profile of an array of 9 nanoparticles (diameter 50 nm) spaced by 5 nm; the total array length (490 nm) is close to double the waist of the simulated Gaussian (500 nm). The horizontal dotted lines represent the estimated signal contrast between regions with and without nanoparticles.}
  \label{Fig22}
\end{figure}

We can convolve the calculated point spread functions of Fig.\ 4\mybox{\textbf{c}} with the (one-dimensional) radial profile of a linear array of nanoparticles (nominal diameter 50 nm, assumed to be spaced by 5 nm) in order to obtain the simulated cathodoluminescence profile; this is shown in Suppl.\ fig.\ \ref{Fig22}. 

It is clear that the modulation depth of luminescence from individual nanoparticles goes from 38 $\%$ at 2 keV to 17 $\%$ at 5 and 10 keV. Nevertheless, a diffraction-limited optical Gaussian beam, taken to have a 250 nm waist, performs even more poorly, as the luminescence in the presence of the nanoparticle array is not modulated at all.

\clearpage

\clearpage

\providecommand{\latin}[1]{#1}
\providecommand*\mcitethebibliography{\thebibliography}
\csname @ifundefined\endcsname{endmcitethebibliography}
  {\let\endmcitethebibliography\endthebibliography}{}